\documentclass[12pt]{article}
\usepackage{amsmath}
\usepackage{amsfonts}
\usepackage{amssymb}
\usepackage{graphicx}
\newtheorem{theorem}{Theorem}

\newtheorem{conjecture}[theorem]{Conjecture}
\newtheorem{corollary}[theorem]{Corollary}

\newtheorem{proposition}[theorem]{Proposition}

\newenvironment{proof}[1][Proof]{\textbf{#1.} }{\ \rule{0.5em}{0.5em}}
\newdimen\dummy
\dummy=\oddsidemargin
\begin{document}

\title{Wulff construction in statistical mechanics and in combinatorics}
\author{Senya Shlosman\thanks{The work was partially supported by the Russian Fund for
Fundamental research through grant 99-01-00284.}\\Centre de Physique Theorique, \\CNRS, Luminy, case 907, F-13288, \\Marseille Cedex 9, France, and \\IPPI, RAS, Moscow, Russia\\\textit{shlosman@cpt.univ-mrs.fr}}
\maketitle
\begin{abstract}
We present the geometric solutions to some variational problems of statistical
mechanics and combinatorics. Together with the Wulff construction, which
predicts the shape of the crystals, we discuss the construction which exhibit
the shape of a typical Young diagram and of a typical skyscraper.

\textbf{Key words and phrases:} Wulff construction, Ising model, droplets,
metastability, facet, Young diagram, hook formula, MacMahon formula.
\end{abstract}

\section{\smallskip Introduction}

In this review we will discuss the phenomenological geometrical construction,
called Wulff construction. Its importance lies in the fact that its
predictions turn out to be true theorems of mathematical physics and
combinatorics. Below we present some of the rigorous results in which this
construction yields the answer to important questions.

\subsection{Statistical mechanics}

The variational problems of statistical mechanics we are going to discuss here
are those related to the formation of a droplet or a crystal of one substance
inside another. The question here is: what shape such a formation would take?
The statement that such shape should be defined by the minimum of the overall
surface energy subject to the volume constraint was known from the times
immemorial. In the isotropic case, when the surface tension does not depend on
the orientation of the surface, and so is just a positive number, the shape in
question should be of course spherical (provided we neglect the gravitational
effects). In a more general situation the shape in question is less symmetric.
The corresponding variational problem is called the \textit{Wulff problem}.
Wulff formulated it in his paper \cite{W} of 1901, where he also presented a
geometric solution to it, called the \textit{Wulff construction} (see section
2.2 below).

This Wulff construction was considered by the rigorous statistical mechanics
as just a phenomenological statement, though the notion of the surface tension
was among its central notions. The situation changed after the appearance of
the book \cite{DKS}. There it was shown that in the setting of the canonical
ensemble formalism, in the regime of the first order phase transition, the
(random) shape occupied by one of the phases has asymptotically (in the
thermodynamic limit) a non-random shape, given precisely by the Wulff
construction! In other words, a typical macroscopic random droplet looks very
close to the Wulff shape. The results of the book \cite{DKS} are restricted to
the 2D Ising ferromagnet at low temperature, though the methods of the book
are suitable for the rigorous treatment of much more general models.

Another problem where Wulff construction is of importance is the theory of
metastability. This is a theory which studies the states of the matter which
``should not be there'', but which still can be observed, albeit for only a
short time. One example is water cooled below the zero temperature. This
supercool water can stay liquid, but not for a long time, and it then freezes
abruptly. Such states are called metastable. They should be thought about as
continuations of the equilibrium states through the point of phase transition,
at which these states cease to exist as equilibrium states. It turns out that
their lifetime is again determined by the quantities given by the Wulff construction.

\subsection{\label{XX} Combinatorics.}

Combinatorics is another field where similar problems appear. Specifically we
have in mind the problem of finding the asymptotic shape of a typical Young
diagram and related problems. This question was first asked and answered by
Vershik and his coauthors, see \cite{VKer,V1,V2, DVZ}. These papers were
treating the two-dimensional case, while the higher dimensional problems were
solved only recently, see \cite{Ke}. In combinatorics there is also a
variational problem to be solved, which in fact is close to the Wulff problem.

When compared with the situation in statistical mechanics, the combinatorial
program and its development look very similar. The only difference is that the
combinatorial counterpart of the Wulff construction was not known, probably
because there was no heuristic period there. Below we describe such a
construction, presented in \cite{S2}. It provides, like the Wulff one, the
geometric solution to the corresponding variational problem under minimal
restrictions on the initial data, and also proves the uniqueness of the solution.\medskip

The next section contains the necessary geometric statements. Some are given
with proofs. In Section 3 we discuss the problems from statistical mechanics,
in which these geometric problems appear naturally. The last Section 4 deals
with the corresponding combinatorial problems.

\section{Geometry}

In this section we consider various geometrical problems. We explain the Wulff
minimization problem and the Wulff construction (\ref{42}), which solves it.
We then describe a related maximization problem. We finish by discussing flat
facets of the Wulff shapes.

\subsection{The droplet energy functional}

Let $S^{d}\subset\mathbb{R}^{d+1}$ denote the unit sphere, and let the real
function $\tau$ on $S^{d}$ be given. We suppose that the function is
continuous, positive: $\tau\left(  \mathbf{\cdot}\right)  \ge const>0,$ and
even: $\tau\left(  \mathbf{n}\right)  =\tau\left(  -\mathbf{n}\right)  .$ Then
for every hypersurface $M^{d}$ embedded in $\mathbb{R}^{d+1}$ we can define
the \textbf{Wulff functional}
\begin{equation}
\mathcal{W}_{\tau}\left(  M^{d}\right)  =\int_{M^{d}}\tau\left(
\mathbf{n}_{x}\right)  \,ds_{x}.\label{01}%
\end{equation}
Here $x\in M^{d}$ is a point on the manifold $M^{d},$ the vector
$\mathbf{n}_{x}$ is the unit vector parallel to the normal to $M^{d}$ at $x,$
and $ds$ is the usual volume $d$-form on $M^{d},$ induced from the Riemannian
metric on $\mathbb{R}^{d+1}$ by the embedding $M^{d}\subset\mathbb{R}^{d+1}.$
Of course, we need to assume that the normal to $M^{d}$ is defined almost
everywhere, i.e. that $M^{d}$ is smooth enough.

Suppose additionally that the hypersurface $M^{d}$ is bounded, closed and is
thus a boundary of a compact region $N^{d+1}\subset\mathbb{R}^{d+1};$ we
denote the volume $\left|  N^{d+1}\right|  $ of $N^{d+1}$ by $\mathrm{vol}%
\left(  M^{d}\right)  ,$ and will call it \textit{the volume inside }$M^{d}.$
We will denote by $D$ the collection of all bounded, closed hypersurfaces
$M^{d} $, \textit{embedded }in $\mathbb{R}^{d+1}.$ We will not distinguish two
elements of $D$ if they are different only by translation. We add to $D$ an
element $\star,$ corresponding to a manifold degenerating to a single point in
$\mathbb{R}^{d+1}.$

Let $h>0$ be a real number. Consider the \textbf{droplet energy functional}
$\Phi$ on $D,$ defined by
\begin{equation}
\Phi\left(  M^{d}\right)  =\mathcal{W}_{\tau}\left(  M^{d}\right)
-h\,\mathrm{vol}\left(  M^{d}\right)  .\label{41}%
\end{equation}

The physical meanings of these functionals are the following: we think about
the elements of $D$ as droplet shapes, and the functional $\mathcal{W}_{\tau}$
represents the surface energy of a droplet. In the regime of droplet
condensation the energy $\Phi$ of a droplet consists of two parts, and in
addition to the surface term we have also the volume term, which encourages
the growth of the droplet.

\smallskip This functional has a local minimum at the point $M_{\min}%
^{d}=\star,$ when the droplet $M^{d}$ degenerates into a single point, with
$\Phi(\star)=0.$ For small droplets $M^{d}$ it is positive, and it takes
negative values for very large $M$-s, when the volume term dominates over the
surface term. Therefore $\Phi$ has to have saddle points, and these saddle
points are of importance for us if we want to understand the process of
droplet condensation. The following theorem describes the structure of the
critical points of $\Phi.$ It turns out to be surprisingly simple!

\begin{theorem}
\smallskip\label{fi} Let the functional $\Phi$ be given by (\ref{41}), with
$h>0$ and the function $\tau$ satisfying the properties given above. Then
$\Phi$ has precisely \textbf{two critical points} in $D.$ One is a local
minimum at $M_{\min}^{d}=\star.$ The other one is a saddle point
\begin{equation}
M_{\mathrm{sdl}}^{d}=\frac{d}{h}\cdot W_{\tau}.\label{72}%
\end{equation}
Here $\frac{d}{h}$ is a dilatation factor, while $W_{\tau}$ is the
\textbf{Wulff shape}, defined by
\begin{equation}
W_{\tau}=\partial K_{\tau}^{<};\text{ with }K_{\tau}^{<}=\left\{
\mathbf{x}\in\mathbb{R}^{d+1}:\forall\mathbf{n\;}\left(  \mathbf{x}%
,\mathbf{n}\right)  \leq\,\tau\left(  \mathbf{n}\right)  \right\}  .\label{42}%
\end{equation}
The value $\Phi\left(  M_{\mathrm{sdl}}^{d}\right)  $ is given by
\begin{equation}
\Phi\left(  M_{\mathrm{sdl}}^{d}\right)  =\frac{1}{d+1}\left(  \frac{d}%
{h}\right)  ^{d}\mathcal{W}_{\tau}\left(  \,W_{\tau}\right)  .\label{73}%
\end{equation}
\end{theorem}

Note that the Wulff body $K_{\tau}^{<}$ is a symmetric convex bounded region
in $\mathbb{R}^{d+1}.$ If at some boundary point the region $K_{\tau}^{<}$ has
a unique support plane, then this plane is of the form
\[
L_{\tau}\left(  \mathbf{n}\right)  =\left\{  \mathbf{x}\in\mathbb{R}%
^{d+1}:\left(  \mathbf{x},\mathbf{n}\right)  =\tau\left(  \mathbf{n}\right)
\right\}
\]
for the corresponding $\mathbf{n.}$ Therefore
\begin{equation}
\mathrm{vol}\left(  W_{\tau}\right)  =\frac1{d+1}\mathcal{W}_{\tau}\left(
W_{\tau}\right)  .\label{46}%
\end{equation}
With this remark the Theorem \ref{fi} follows easily from the Theorem \ref{w}
of the next subsection.

\subsection{Wulff minimizing problem.}

Let now $D_{q}$ be the collection of all closed hypersurfaces $M^{d}$,
embedded\textit{\ }in $\mathbb{R}^{d+1},$ and such that the volume
$\mathrm{vol}\left(  M^{d}\right)  $ inside\textit{\ }$M^{d}$ equals $q$. The
\textit{Wulff problem }consists in finding the lower bound of $\mathcal{W}%
_{\tau}$ over $D_{q}:$%
\begin{equation}
w_{\tau}=\inf_{M\in D_{q}}\mathcal{W}_{\tau}\left(  M\right)  ,\label{02}%
\end{equation}
as well as the minimizing surface(s) $W_{\tau}^{\left(  q\right)  },$ such
that $\mathcal{W}_{\tau}\left(  W_{\tau}^{\left(  q\right)  }\right)
=w_{\tau}^{\left(  q\right)  },$ if it exists.

\begin{theorem}
\label{w} The variational problem (\ref{02}) has a unique (up to translations)
solution
\[
W_{\tau}^{\left(  q\right)  }=\left(  \frac{q\left(  d+1\right)  }%
{\mathcal{W}_{\tau}\left(  W_{\tau}\right)  }\right)  ^{1/\left(  d+1\right)
}W_{\tau},
\]
which is a scaled version of the Wulff shape, see (\ref{42}).
\end{theorem}

The paper \cite{T2} contains a simple proof that $\mathcal{W}_{\tau}\left(
W_{\tau}^{\left(  q\right)  }\right)  \le\mathcal{W}_{\tau}\left(  M\right)  $
for every $M\in D_{q}.$ It proceeds in four steps, as follows. Let $M=\partial
N,$ $W_{\tau}=\partial K_{\tau}^{<}.$

\noindent$1)$ We observe first, that
\begin{equation}
\mathcal{W}_{\tau}(M)\ge\lim_{\varepsilon\to0}\frac{|N+\varepsilon K_{\tau
}^{<}|-|N|}\varepsilon.\label{43}%
\end{equation}
Here for $A,B\subset\mathbb{R}^{d+1}$ we denote by $A+B=\left\{
z\in\mathbb{R}^{d+1}:z=x+y,x\in A,y\in B\right\}  .$ To see that (\ref{43})
holds, we note first that by continuity it is enough to check (\ref{43}) in
case when $N$ is a polyhedron. Then (\ref{43}) follows from the remark that
for a hyperplane $L\left(  \mathbf{n}\right)  ,$ orthogonal to $\mathbf{n,}$
the sum $L\left(  \mathbf{n}\right)  +K_{\tau}^{<}$ is a strip around
$L\left(  \mathbf{n}\right)  $ with the width $\le\tau\left(  \mathbf{n}%
\right)  ,$ while the contributions of the lower-dimensional faces of $N$ to
$|N+\varepsilon K_{\tau}^{<}|$ are of the order of at most $\varepsilon^{2}.$

\noindent$2)$ Next, we apply Brunn-Minkowski inequality, which states that
\[
\left|  A+B\right|  \ge(|A|^{1/\left(  d+1\right)  }+|B|^{1/\left(
d+1\right)  })^{d+1}.
\]
It implies that
\begin{align*}
|N+\varepsilon K_{\tau}^{<}|  & \ge(|N|^{1/\left(  d+1\right)  }%
+\varepsilon|K_{\tau}^{<}|^{1/\left(  d+1\right)  })^{\left(  d+1\right)  }\\
& \ge|N|+\left(  d+1\right)  \varepsilon|N|^{d/\left(  d+1\right)  }|K_{\tau
}^{<}|^{1/\left(  d+1\right)  }.
\end{align*}

\noindent$3)$ Without loss of generality we can restrict ourselves to the case
$q=\mathrm{vol}\left(  W_{\tau}\right)  .$ Then we further have
\[
\frac{|N+\varepsilon K_{\tau}^{<}|-|N|}\varepsilon\ge\left(  d+1\right)
\mathrm{vol}\left(  W_{\tau}\right)  .
\]

\noindent$4)$ On the other hand, $\mathcal{W}_{\tau}\left(  W_{\tau}\right)
=\left(  d+1\right)  \mathrm{vol}\left(  W_{\tau}\right)  ,$ see (\ref{46}).
\endproof

The uniqueness of the minimizing surface is proven in \cite{T1}. The proof is
much more involved.

\subsection{\smallskip Two-dimensional case}

In what follows we fix $q=\mathrm{vol}\left(  W_{\tau}\right)  $ and omit it
from our notation. The specifics of dimension 2 is that the minimizer
$W_{\tau}$ of the functional $\mathcal{W}_{\tau}$ is not only unique, but also
is stable in the Hausdorf metric. To formulate that statement we first remind
the reader that the Hausdorf distance $\rho_{H}$ between the two sets
$A,C\in\mathbb{R}^{d}$ is defined as
\[
\rho_{H}\left(  A,C\right)  =\min\left\{  \inf\left[  r:A\subset
C+B_{r}\right]  ,\inf\left[  r:C\subset A+B_{r}\right]  \right\}  ,
\]
where $B_{r}$ is the ball of radius $r.$

\begin{theorem}
\label{d2} Let $\gamma$ be a closed curve in $\mathbb{R}^{2},$ with the area
inside it equals to that of $W_{\tau}.$ Suppose that for $0<\varepsilon<1$
\[
\mathcal{W}_{\tau}\left(  \gamma\right)  \leq\mathcal{W}_{\tau}\left(
W_{\tau}\right)  +\varepsilon.
\]
Then there exists a shift $\tilde{\gamma}$ of $\gamma$ by a vector
$\mathbf{x=x}\left(  \gamma\right)  $ in $\mathbb{R}^{2},$ such that
\[
\rho_{H}\left(  \tilde{\gamma},W_{\tau}\right)  \leq C\sqrt{\varepsilon},
\]
where $C=C\left(  \tau\right)  >0.$
\end{theorem}

Of course, Theorem \ref{d2} implies uniqueness of the minimizing curve of the
functional $\mathcal{W}_{\tau}.$ The Theorem \ref{d2} fails in higher dimensions.

For the proof, see \cite{DKS}, Sect. 2.4.

\subsection{ Maximizing problem.\label{2.4}}

In a dual problem we again have a function $\eta$ of a unit vector, but this
time it is defined only over the subset $\Delta^{d}=$ $S^{d}\cap\mathbb{R}%
_{+}^{d+1}\ $of them, lying in the positive octant. We suppose again that the
function is continuous and nonnegative: $\eta\left(  \mathbf{\cdot}\right)
\ge0.$ We assume additionally that
\begin{equation}
\eta\left(  \mathbf{n}\right)  \rightarrow0\text{ uniformly as }%
\mathbf{n}\rightarrow\partial\,\Delta^{d}.\label{05}%
\end{equation}
Let now $G\subset\mathbb{R}_{+}^{d+1}$ be an embedded hypersurface. We assume
that for almost every $x\mathbf{\in}G$ the normal vector $\mathbf{n}_{x}$ is
defined, and moreover
\begin{equation}
\mathbf{n}_{x}\in\Delta^{d}\text{ for a.e. }x\mathbf{\in}G.\label{07}%
\end{equation}
Then we can define the functional
\begin{equation}
\mathcal{V}_{\eta}\left(  G\right)  =\int_{G}\eta\left(  \mathbf{n}%
_{x}\right)  \,ds_{x}.\label{03}%
\end{equation}
In analogy with the Section 2.2 we introduce the families $\bar{D}_{q},q>0,$
of such surfaces $G$ as follows:

$G\in\bar{D}_{q}$ iff

\noindent$\,i)$ $\,G$ splits the octant $\mathbb{R}_{+}^{d+1}$ in two parts,
with the boundary $\partial\mathbb{R}_{+}^{d+1}$ belonging to one of them,

\noindent$ii)$ the ($\left(  d+1\right)  $-dimensional) volume of the body
$Q\left(  G\right)  ,$ enclosed between $\partial\mathbb{R}_{+}^{d+1}$ and
$G,$ equals $q.$ In what follows we denote the volume of $Q\left(  G\right)  $
by $\mathrm{vol}\left(  G\right)  .$

\noindent For example, let $f\left(  y\right)  \ge0$ be a function on
$\mathbb{R}_{+}^{d},$ non-increasing in each of $d$ variables, and $G\left[
f\right]  \subset$ $\mathbb{R}_{+}^{d+1}$ be its graph. Then
\begin{equation}
\mathrm{vol}\left(  G\left[  f\right]  \right)  =\int_{\mathbb{R}_{+}^{d}%
}f\left(  y\right)  \,dy,\label{06}%
\end{equation}
so if $\int_{\mathbb{R}_{+}^{d}}f\left(  y\right)  \,dy=q,$ then $G\left[
f\right]  $ is an element of $\bar{D}_{q},$ provided the function $f$ is
sufficiently smooth.

Our problem now\textit{\ }is to find the \textit{upper bound} of
$\mathcal{V}_{\eta}$ over $\bar{D}_{1}:$%
\begin{equation}
v_{\eta}=\sup_{G\in\bar{D}_{1}}\mathcal{V}_{\eta}\left(  G\right)  ,\label{04}%
\end{equation}
as well as the maximizing surface(s) $V_{\eta}\in\bar{D}_{1},$ such that
$\mathcal{V}_{\eta}\left(  V_{\eta}\right)  =v_{\eta},$ if the maximizer does
exist. Note that the last problem differs crucially from (\ref{02}), since
here we are looking for the \textit{supremum. }In particular, this upper bound
evidently diverges if taken over all surfaces, and not only over ``monotone''
one, in the sense of (\ref{07}), unlike in the problem (\ref{02}).

It turns out that there exists a geometric construction, which provides a
solution to the variational problem (\ref{04}), in the same way as the Wulff
construction solves the problem (\ref{02}). It was found in \cite{S2}.

Let
\begin{equation}
K_{\eta}^{>}=\left\{  \mathbf{x}\in\mathbb{R}^{d+1}:\forall\mathbf{n}\in
\Delta^{d}\;\left(  \mathbf{x},\mathbf{n}\right)  \ge\,\eta\left(
\mathbf{n}\right)  \right\}  ,\label{15}%
\end{equation}%
\begin{equation}
G_{\eta}=\partial\left(  K_{\eta}^{>}\right)  .\label{16}%
\end{equation}
Because of (\ref{05}), the surface $G_{\eta}$ is a graph of a function,
$f_{\eta}\left(  y\right)  ,\,y\in\mathbb{R}_{+}^{d},$ i.e. $G_{\eta}=G\left[
f_{\eta}\right]  .$ Note, that the integral $\mathrm{vol}\left(  G_{\eta
}\right)  $ (see (\ref{06})) satisfies
\begin{equation}
\mathrm{vol}\left(  G_{\eta}\right)  =\frac{\mathcal{V}_{\eta}\left(  G_{\eta
}\right)  }{d+1}\label{75}%
\end{equation}
(compare with (\ref{46})), provided both sides of (\ref{75}) are finite.

\begin{theorem}
\label{dw} Suppose the integral $\mathrm{vol}\left(  G_{\eta}\right)  $ is
converging. Then the functional $\mathcal{V}_{\eta}$ has a unique maximizer,
$V_{\eta},$ over the set $\bar{D}_{1}.$ It is given by the dilatation of the
surface (\ref{16}):
\begin{equation}
V_{\eta}=\left(  \frac{d+1}{\mathcal{V}_{\eta}\left(  G_{\eta}\right)
}\right)  ^{1/\left(  d+1\right)  }G_{\eta}.\label{18}%
\end{equation}
If the integral $\mathrm{vol}\left(  G\right)  $ diverges, then $v_{\eta
}=\infty.$
\end{theorem}

\subsection{Facets}

In this subsection we will discuss facets -- the flat pieces of the surfaces
$W_{\tau}$ and $V_{\eta}.$

In general, different functions $\tau$ can result in the same Wulff shape
$W_{\tau}$. There is a way to make this correspondence one-to one, by
requiring additionally that the function $\tau$ satisfies the pyramid
inequality, introduced in \cite{DS}. Let us formulate it for the
two-dimensional case. Let $A,B,C$ be three points in $\mathbb{R}^{2}.$
Consider three unit vectors $\mathbf{n}_{AB},\mathbf{n}_{BC},\mathbf{n}_{CA},$
orthogonal to the respective sides of the triangle $ABC.$ The pyramid (or in
this case triangle) inequality is the requirement that
\[
\left|  AB\right|  \tau\left(  \mathbf{n}_{AB}\right)  +\left|  BC\right|
\tau\left(  \mathbf{n}_{BC}\right)  \ge\left|  CA\right|  \tau\left(
\mathbf{n}_{CA}\right)
\]
for every triangle $ABC.$ The pyramid inequality is equivalent to the property
that every plane
\[
L_{\tau}^{=}\left(  \mathbf{n;}\lambda\right)  =\left\{  \mathbf{x}%
\in\mathbb{R}^{d+1}:\left(  \mathbf{x},\mathbf{n}\right)  =\lambda
\,\tau\left(  \mathbf{n}\right)  \right\}
\]
is a support plane to the (convex) body $K_{\tau}^{<}\left(  \lambda\right)  $
(see (\ref{42})). In the case when $\tau$ is the true surface tension function
of some statistical mechanical system, this property is expected always to
hold. In case of the 2D Ising model the triangle inequality holds sharply, see
\cite{DKS}. Sharp pyramid inequality implies smoothness of the corresponding
Wulff shape. For some general class of models the pyramid inequality was shown
to hold in \cite{Messager}. In what follows we will suppose it to hold. The
validity of the pyramid inequality is equivalent to the convexity of the
affine extension $\hat{\tau}$ of $\tau$ to $\mathbb{R}^{d+1}$ defined by
$\hat{\tau}\left(  \mathbf{x}\right)  =\left|  \left|  \mathbf{x}\right|
\right|  \tau\left(  \frac{\mathbf{x}}{\left|  \left|  \mathbf{x}\right|
\right|  }\right)  .$ As a result, the function $\tau$ is differentiable a.e.,
and it has directional derivatives at any point $\mathbf{n}\in S^{d}$ along
any tangent direction. We will denote by $\tau^{\prime}\left(  \mathbf{\nu
}_{\mathbf{n}}\right)  $ the derivative of the function $\tau$ taken at the
point $\mathbf{n}$ along the unit tangent vector $\mathbf{\nu}_{\mathbf{n}}$
to $S^{d}$ at $\mathbf{n}.$ The set of all such unit tangent vectors will be
denoted by $S_{\mathbf{n}}^{d-1}.$

We say that the function $\tau$ has a cusp at the point $\mathbf{n}_{0}$ if
for every\textbf{\ }$\mathbf{\nu\in}S_{\mathbf{n}_{0}}^{d-1}$ we have
\begin{equation}
\tau^{\prime}\left(  \mathbf{\nu}\right)  \neq-\tau^{\prime}\left(
-\mathbf{\nu}\right)  .\label{30}%
\end{equation}
(Note that at any point of smoothness of $\tau$ we have $\tau^{\prime}\left(
\mathbf{\nu}\right)  =-\tau^{\prime}\left(  -\mathbf{\nu}\right)  .$) If that
is the case, then the intersection of the support plane $L_{\tau}^{=}\left(
\mathbf{n}_{0}\mathbf{;}\lambda\right)  $ with the Wulff body $K_{\tau}%
^{<}\left(  \lambda\right)  $ is a convex closed subset of this plane, with
nonempty interior. Such flat piece of the boundary $\partial\left(  K_{\tau
}^{<}\left(  \lambda\right)  \right)  $ is called a \textbf{facet}. The shape
of the facet can then be found by an analog of the Wulff construction
(\ref{42}), applied to the function $\tau^{\prime}\left(  \mathbf{\nu}\right)
$ on $S_{\mathbf{n}_{0}}^{d-1}.$ This analog is especially simple if there is
an additional symmetry of the function $\tau$ at $\mathbf{n}_{0}$ - namely,
that for every $\mathbf{\nu}\in S_{\mathbf{n}_{0}}^{d-1}$ we have
$\tau^{\prime}\left(  \mathbf{\nu}\right)  =\tau^{\prime}\left(  -\mathbf{\nu
}\right)  .$ Note that (\ref{30}) then implies that $\tau^{\prime}\left(
\mathbf{\nu}\right)  >0$ for every $\mathbf{\nu}\in S_{\mathbf{n}_{0}}^{d-1}.$
In this case the above mentioned analog is just the Wulff construction itself,
see \cite{Miracle}. Namely, the facet shape is given by
\begin{equation}
F\left(  \tau,\mathbf{n}_{0}\right)  =\left\{  \mathbf{x}\in\mathbb{R}%
^{d}:\forall\mathbf{\nu\;}\left(  \mathbf{x},\mathbf{\nu}\right)  \le
\,\tau^{\prime}\left(  \mathbf{\nu}\right)  \right\}  .\label{50}%
\end{equation}

All said above is applicable to facets of the surfaces $V_{\eta}.$ The only
difference is that in addition to bounded facets they may have unbounded ones,
lying in ortants $\mathbb{R}_{+}^{d}$ $\subset\partial\mathbb{R}_{+}^{d+1}.$
The shape of the bounded facets is given by the straightforward analog of
(\ref{50}). The unbounded facet in, say, $xy$ plane (in $d+1=3$-dimensional
case) appears if the function $\eta$ has nonzero derivatives $\eta^{\prime
}\left(  \nu\right)  $ at the point $\mathbf{n}_{z}=(0,0,1)$ along tangent
directions $\nu\in\left[  0,\frac\pi2\right]  \equiv\Delta^{1}$ to the
spherical triangle $\Delta^{2}=S^{2}\cap\mathbb{R}_{+}^{3}$. In such a case
the facet shape is given by the analog of (\ref{15}):
\begin{equation}
F\left(  \eta,\mathbf{n}_{z}\right)  =\left\{  \mathbf{x}\in\mathbb{R}%
^{2}:\forall\mathbf{\nu\;}\left(  \mathbf{x},\mathbf{\nu}\right)  \ge
\,\eta^{\prime}\left(  \mathbf{\nu}\right)  \right\}  .\label{83}%
\end{equation}

\section{Statistical mechanics}

The relevance of the Wulff construction for the rigorous statistical mechanics
lies in the fact that it describes the asymptotic shape of the droplet of one
phase on the background of another phase in the regime of phase coexistence.
The first result of that kind was obtained in the book \cite{DKS}. In what
follows, we briefly summarize it, as well as more recent results. We will talk
only about the Ising model, though more general models can be studies as well.

The rest of this section is organized as follows. We first recall some basic
facts about the Ising model. We then formulate the results about the shape of
the droplets of the $\left(  +\right)  $-phase in the $\left(  -\right)
$-phase, in the two-dimensional case and in higher dimensions. We conclude by
explaining the relevance of these results to the theory of metastability.

\subsection{Basics about the Ising model}

\textit{Gibbs states: }At each site in $\mathbb{Z}^{d+1}$ there is a spin
which can take values $-1$ and $+1$. The configurations will therefore be
elements of the set $\{-1,+1\}^{\mathbb{Z}^{d+1}}=\Omega$. Given $\sigma
\in\Omega$, we write $\sigma(x)$ for the spin at the site $x\in\mathbb{Z}%
^{d+1}$. Two configurations are specially relevant, the one with all spins
$-1$ and the one with all spins $+1$. We will use the simple notation $-$ and
$+$ to denote them.

Observables are just functions on $\Omega$. Local observables are those which
depend only on the values of finitely many spins.

We will consider the formal Hamiltonian
\begin{equation}
H_{h}(\sigma)=-\frac12\sum_{x,y\text{ n.n.}}\sigma(x)\sigma(y)-\frac
h2\sum_{x}\sigma(x),\label{70}%
\end{equation}
where $h\in\mathbb{R}^{1}$ is the external field and $\sigma\in\Omega$ is a
generic configuration. We define, for each set $\Lambda\subset\subset
\mathbb{Z}^{d+1}$ and each \textit{boundary condition} $\xi\in\Omega$,
\[
H_{\Lambda,\xi,h}(\sigma)=-\frac12\sum_{\substack{x,y\text{ n.n.}
\\x,y\in\Lambda}}\sigma(x)\sigma(y)-\frac12\sum_{\substack{x,y\text{ n.n.}
\\x\in\Lambda,y\not \in\Lambda}}\sigma(x)\xi(y)-\frac h2\sum_{x\in\Lambda
}\sigma(x).
\]
We introduce the subset $\Omega_{\Lambda,\xi}\subset\Omega$ as $\Omega
_{\Lambda,\xi}=\left\{  \sigma\in\Omega:\sigma\left(  x\right)  =\xi\left(
x\right)  \text{ for }x\notin\Lambda\right\}  .$ We further define the
partition function
\begin{equation}
Z_{\Lambda,\xi,T,h}=\sum_{\sigma\in\Omega_{\Lambda,\xi}}\exp(-\beta
H_{\Lambda,\xi,h}(\sigma)),\label{52}%
\end{equation}
where $\beta=1/T$.

The \textit{Grand Canonical Gibbs measure} in $\Lambda$ with boundary
condition $\xi$ under external field $h$ and at temperature $T$ is defined on
$\Omega$ as
\begin{equation}
\mu_{\Lambda,\xi,T,h}(\sigma)=\left\{
\begin{array}
[c]{ll}%
Z_{\Lambda,\xi,T,h}^{-1}\exp(-\beta H_{\Lambda,\xi,h}(\sigma)) & \text{if
$\sigma\in\Omega_{\Lambda,\xi}$},\newline \text{ \quad}\\
0 & \text{otherwise.}%
\end{array}
\right. \label{51}%
\end{equation}

For each value of $T$ and $h$, the Gibbs measure $\mu_{\Lambda(l),\pm,T,h} $
with $\left(  \pm\right)  $-boundary conditions in the square box $\Lambda(l)$
of size $l$ converges weakly, as $l\rightarrow\infty$, to a probability
measure that we will denote by $\mu_{\pm,T,h}$. If $h\neq0,$ then $\mu
_{-,T,h}=\mu_{+,T,h}$, so it will be denoted simply by $\mu_{T,h} $. If $h=0$
the same is true if the temperature is larger than or equal to a critical
value $T_{c}=T_{c}\left(  d+1\right)  $, and is false for $T<T_{c}$, in which
case one says that there is phase coexistence. The critical temperature
depends on dimension; $T_{c}\left(  1\right)  =0,$ while $T_{c}\left(
d+1\right)  >0$ for $d\ge1.$ The measure $\mu_{+,T,0}\equiv\mu_{+,T}$ is
called the\textit{\ }$(+)$-phase\textit{, }and\textit{\ }$\mu_{-,T}$ -- the
$\left(  -\right)  $-phase.

In case $\Lambda\ $is a cubic box, $\Lambda(l),$ we can also talk about
\textit{periodic boundary conditions} in $\Lambda,$ which correspond to
wrapping $\Lambda(l)$ into a torus. Then we can similarly define the
configuration spaces $\Omega_{\Lambda(l)}=\{-1,+1\}^{\Lambda(l)},$ the
Hamiltonian $H_{\Lambda(l),per,h}(\sigma)$ for $\sigma\in\Omega_{\Lambda(l)},$
the partition function $Z_{\Lambda(l),per,T,h}$ and the corresponding Grand
Canonical Gibbs measure $\mu_{\Lambda(l),per,T,h}$ in $\Lambda\left(
l\right)  $ -- with periodic boundary condition under external field $h$ and
at temperature $T.$

Next thing we need to introduce is the notion of the \textit{Canonical Gibbs
measures}. To do it we first fix a real number $\rho,$ $-1<\rho<1.$ Now, for
every finite box $\Lambda$ we consider the subset $\Omega_{\Lambda,\xi}^{\rho
}\subset\Omega_{\Lambda,\xi},$ defined by the property:
\[
\sigma\in\Omega_{\Lambda,\xi}^{\rho}\Longleftrightarrow\sum_{x\in\Lambda
}\sigma\left(  x\right)  =\left[  \rho\left|  \Lambda\right|  \right]  ,
\]
where $\left[  \cdot\right]  $ stays for the integer part. The canonical
partition function $Z_{\Lambda,\xi,T,h}^{\rho}$ is defined by the analog of
(\ref{52}), where the summation range $\Omega_{\Lambda,\xi}$ is restricted to
$\Omega_{\Lambda,\xi}^{\rho}.$ The Canonical Gibbs measure $\mu_{\Lambda
,\xi,T,h}^{\rho}$ is defined by the analog of (\ref{51}), where $Z_{\Lambda
,\xi,T,h}$ is replaced by $Z_{\Lambda,\xi,T,h}^{\rho},$ and $\Omega
_{\Lambda,\xi}$ - by $\Omega_{\Lambda,\xi}^{\rho}.$ The measures $\mu
_{\Lambda(l),per,T,h}^{\rho}$ are introduced in the obvious way. One checks
easily that the measures $\mu_{\Lambda,\xi,T,h}^{\rho},\,\mu_{\Lambda
(l),per,T,h}^{\rho}$ in fact do not depend on the magnetic field $h,$ so we
will denote them by $\mu_{\Lambda(l),per,T}^{\rho},$ $\mu_{\Lambda,\xi
,T}^{\rho}.$

One can say that canonical Gibbs measures are obtained from the grand
canonical one by conditioning according to the event $\sum_{x\in\Lambda}%
\sigma\left(  x\right)  =\left[  \rho\left|  \Lambda\right|  \right]  .$
Another collection of states we will need is obtained from the grand canonical
Gibbs distribution by conditioning over the event $\left|  \sum_{x\in\Lambda
}\sigma\left(  x\right)  \right|  \le\left[  \rho\left|  \Lambda\right|
\right]  .$ They will be denoted by $\mu_{\Lambda,\xi,T,h}^{\le\rho}.$
Alternatively one can introduce the set $\Omega_{\Lambda,\xi}^{\le\rho}$ by
considering only the configurations $\sigma$ with $\left|  \sum_{x\in\Lambda
}\sigma\left(  x\right)  \right|  \le\left[  \rho\left|  \Lambda\right|
\right]  ,$ the corresponding partition function $Z_{\Lambda,\xi,T,h}^{\le
\rho}$ and then again write the analog of (\ref{51}).

For the expected value corresponding to a Gibbs measure $\mu_{{...}}^{{...}} $
we will use the notation $\langle f\rangle_{{...}}^{{...}}=\int fd\mu_{{...}%
}^{{...}},$ where ${...}$ stands for arbitrary sub- and superscripts. The
spontaneous magnetization at temperature $T$ is defined as $m^{*}%
(T)=\langle\sigma\left(  0\right)  \rangle_{+,T}.$ It is known that
$m^{*}(T)>0$ if and only if $T<T_{c}$.

\textit{Contours. }With every configuration $\sigma$ one can associate a
collection $\gamma\left(  \sigma\right)  $ of surfaces in $\mathbb{R}^{d+1},$
which are called contours. They are defined in the following way. Let us
consider the set $\frak{B}\left(  \sigma\right)  $ of all lattice bonds $x,y$
such that $\sigma\left(  x\right)  \sigma\left(  y\right)  =-1.$ The surfaces
in the collection $\gamma\left(  \sigma\right)  $ are formed by the collection
$\frak{P}\left(  \sigma\right)  $ of $d$-dimensional plaquettes, which are
dual to the bonds in $\frak{B}\left(  \sigma\right)  .$ Each connected
component of $\frak{P}\left(  \sigma\right)  $ is called a contour.

\textit{Surface tension: }The direction dependent surface tension is defined
in the following way. For each vector $\mathbf{n}\in S^{d}$, consider the
configuration $\xi(\mathbf{n})$, to be used as a boundary condition:
\[
\xi(\mathbf{n})(x)=\left\{
\begin{array}
[c]{ll}%
+1, & \text{if $(x,\mathbf{n})\geq0$},\\
-1, & \newline \text{if $(x,\mathbf{n})<0,$}%
\end{array}
\right.
\]
where $\left(  \cdot,\cdot\right)  $ is the usual scalar product in
$\mathbb{R}^{d+1}$. The surface tension along the plane perpendicular to
$\mathbf{n}$ is given by
\begin{equation}
\tau_{T}(\mathbf{n})=\lim_{l\rightarrow\infty}-\frac1{\beta A\left(
l,\mathbf{n}\right)  }\log\frac{Z_{\Lambda(l),\xi(\mathbf{n}),T,0}}%
{Z_{\Lambda(l),+,T,0}},\label{55}%
\end{equation}
where $A\left(  l,\mathbf{n}\right)  \ $is the ($d$-dimensional) area of the
intersection of the hyperplane $\left\{  \mathbf{x:}(\mathbf{x},\mathbf{n}%
)=0\right\}  $ with the box $\Lambda(l)$. It is known that for each $T<T_{c}$
the surface tension $\tau_{T}(\cdot)$ is a continuous strictly positive and
finite function.

\subsection{``The big droplet has the Wulff shape''. Dimension $2$.}

The measure $\mu_{\Lambda(l),per,T}^{\rho}$ on configurations can be viewed as
measures on contours. The theorem below explains what are the typical
properties of these random contours, which explanation validates the title of
the present subsection.

In the theorem below we consider the values of the density $\rho$ smaller than
the average value $\langle\sigma\left(  0\right)  \rangle_{\Lambda
(l),per,T}\approx m^{*}(T).$ That forces the extra amount of the $\left(
-\right)  $-phase in the $\left(  +\right)  $-phase. As the reader expects
perhaps, this extra amount gathers into one big droplet, which has
approximately the Wulff shape.

The rest of this subsection deals exclusively with the case of dimension 2.

\begin{theorem}
\label{w2} Let $T<T_{c}\left(  2\right)  .$ Then for every density $\rho$ with
$m^{\ast}(T)/2<\left|  \rho\right|  <m^{\ast}(T)$ there exist a sequence
$\frak{D}_{l}$ of subsets of configurations, $\frak{D}_{l}\subset
\Omega_{\Lambda(l)}^{\rho},$ which are \textbf{typical}, that is
\[
\mu_{\Lambda(l),per,T}^{\rho}\left(  \frak{D}_{l}\right)  \rightarrow1\text{
as }l\rightarrow\infty,
\]
and which have the following properties:

\begin{itemize}
\item  for every configuration $\sigma\in\frak{D}_{l}$ the set $\gamma\left(
\sigma\right)  $ of its contours has precisely one ``long'' contour,
$\Gamma\left(  \sigma\right)  \in\gamma\left(  \sigma\right)  ;$ all other
contours are not longer than $K\ln l,$ $K=K\left(  \beta,\rho\right)  ,$

\item  the area $\left|  \mathrm{Int}\,\Gamma\left(  \sigma\right)  \right|  $
inside $\Gamma\left(  \sigma\right)  $ satisfies
\[
\left|  \,\left|  \mathrm{Int}\,\Gamma\left(  \sigma\right)  \right|  -\lambda
l^{2}\right|  \leq Kl^{6/5}\left(  \ln l\right)  ^{\varkappa},
\]
where $\lambda=\frac{m^{\ast}(T)-\left|  \rho\right|  }{2m^{\ast}%
(T)},\varkappa=\varkappa\left(  \rho\right)  ,$

\item  there is a point $x=x\left(  \sigma\right)  \in\Lambda(l),$ such that
the shift of $\Gamma\left(  \sigma\right)  $ by $x\left(  \sigma\right)  $
brings the contour $\Gamma\left(  \sigma\right)  $ very close to the scaled
Wulff curve, defined by the Ising model surface tension $\tau$ (see
\ref{55}):
\begin{equation}
\rho_{H}\left(  \Gamma\left(  \sigma\right)  +x\left(  \sigma\right)
,\sqrt{\frac{2\lambda}{\mathcal{W}_{\tau}\left(  W_{\tau}\right)  }}lW_{\tau
}\right)  \leq Kl^{2/3}\left(  \ln l\right)  ^{\varkappa}.\label{57}%
\end{equation}
\end{itemize}
\end{theorem}

The first version of the above theorem was obtained in \cite{DKS}. The
extension to all subcritical temperatures is due to Ioffe and
Ioffe\&Schonmann, \cite{I1, I2, IS}. The value $2/3$ of the exponent in
(\ref{57}) is an improvement of the original $3/4$ result, and is due to
Alexander \cite{A}.

{\small The restriction }$m^{*}(T)/2<\left|  \rho\right|  ${\small \ in the
theorem is needed because without it the droplet may prefer to shape itself
into a strip between two meridians rather than to take the Wulff shape, see
\cite{S1}. This strip shape is, however, disadvantageous once the volume of
the droplet has to be small.}

\subsection{``The big droplet has the Wulff shape''. Dimension $\ge3$.}

In higher dimensions the status of the analog of the Theorem \ref{w2} is the
following: it might be valid for very low temperatures, and it very probably
fails for intermediate temperatures below the critical temperature.

What is known to hold is a weaker version of this theorem, where the distance
between random droplet and the Wulff shape is measured not by Hausdorf
distance, but is understood in $L^{1}$ sense. To state the corresponding
theorem, we will associate with every configuration $\sigma$ from
$\Omega_{\Lambda(l)}$ a real valued function $M_{\sigma}\left(  t\right)  $ on
the unit torus $\mathbb{T}^{d+1},$ and we then compare this function with the
indicator function $\mathbb{I}_{\lambda K_{\tau}^{<}},$ where $\lambda
K_{\tau}^{<}\subset\mathbb{T}^{d+1}$ is the Wulff body, properly scaled.

The function $M_{\sigma}\left(  t\right)  $ is defined as follows. We denote
by $i_{l}$ the natural embedding of the discrete torus $\Lambda(l)$ into
$\mathbb{T}^{d+1},$ the image of $i_{l}$ being the grid with spacing
$\frac1l.$ For $t\in\mathbb{T}^{d+1}$ we define $b_{l}\left(  t\right)
\subset\mathbb{T}^{d+1}$ to be the ball centered at $t$ with radius
$\sqrt[d+1]{\frac1l},$ and let $B_{l}\left(  t\right)  \subset\Lambda(l)$ be
its preimage under $i_{l}.$ Then
\[
M_{\sigma}\left(  t\right)  =\frac1{\left|  B_{l}\left(  t\right)  \right|
}\sum_{x\in B_{l}\left(  t\right)  }\sigma\left(  x\right)  .
\]
(The choice of the scale $\sqrt[d+1]{\frac1l}$ is somewhat arbitrary here; the
essential requirement is that the size of $B_{l}\left(  t\right)  $ should
grow with $l,$ but stay infinitesimally small compared with $l^{d+1}$.)

We are going to discuss the typical configurations of the measure
$\mu_{\Lambda(l),per,T,h}^{<\rho}$ for $h=0.$ In order to have a droplet,
which, moreover, does not wrap itself around the torus, we have to put
restrictions on $\rho.$ The following one:%

\begin{equation}
m^{*}(T)\left(  1-2\left(  d+1\right)  ^{-\frac{d+1}d}\right)  <\left|
\rho\right|  <m^{*}(T)\label{60}%
\end{equation}
is enough. The straightforward calculation shows that under such restriction
we have to expect to see a droplet $\lambda K_{\tau}^{<}$ with
\begin{equation}
\lambda=\sqrt[d+1]{\frac{d+1}{\mathcal{W}_{\tau}\left(  W_{\tau}\right)  }%
}\frac{m^{*}-\left|  \rho\right|  }{2m^{*}}.\label{61}%
\end{equation}
To make a precise statement we first introduce for every subset $A\subset
\mathbb{T}^{d+1}$ the indicator $\mathbb{I}_{A}\left(  t\right)  =\left\{
\begin{array}
[c]{cc}%
1, & t\in A\\
-1, & t\in A^{c}%
\end{array}
\right.  .$ For every function $v$ in $L^{1}\left(  \mathbb{T}^{d+1}\right)  $
we denote by $U\left(  v,\delta\right)  $ its $\delta$-neighborhood in
$L^{1}\left(  \mathbb{T}^{d+1}\right)  .$

\begin{theorem}
\label{w3} Let the dimension $d+1$ be at least two, and the temperature
$T<T_{c}\left(  d+1\right)  $ satisfies some extra technical restrictions (see
below). Then in the large volume and for densities $\rho$ satisfying
(\ref{60}) the function $M_{\sigma}\left(  t\right)  $ typically has the Wulff
shape: for every $\delta>0$%
\[
\lim_{l\rightarrow\infty}\mu_{\Lambda(l),per,T,0}^{<\rho}\left\{  \frac
{1}{m^{\ast}\left(  T\right)  }M_{\sigma}\left(  \cdot\right)  \in
\bigcup_{t\in\mathbb{T}^{d+1}}\left(  U\left(  \mathbb{I}_{\lambda K_{\tau
}^{<}+t},\delta\right)  \cup U\left(  -\mathbb{I}_{\lambda K_{\tau}^{<}%
+t},\delta\right)  \right)  \right\}  =1.
\]
Here $\lambda$ is given by (\ref{61}).
\end{theorem}

The shifts by all $t$-s of the Wulff shape $\lambda K_{\tau}^{<}$ appear in
the statement since the location of the droplet can be arbitrary. The droplet
$\Gamma\left(  \sigma\right)  $ can be either of $\left(  +\right)  $-phase or
of $\left(  -\right)  $-phase, which is why we consider two indicators:
$\mathbb{I}_{\lambda K_{\tau}^{<}}$ and $-\mathbb{I}_{\lambda K_{\tau}^{<}}.$
Note that if a point $t$ is such that the ball $B_{l}\left(  t\right)  $ stays
away from the boundary of the droplet $\Gamma\left(  \sigma\right)  $ present
in the configuration $\sigma,$ then the value $M_{\sigma}\left(  t\right)  $
should be expected to be $\pm m^{*}\left(  T\right)  ,$ depending on the phase
inside/outside the droplet, which explains the factor $\frac1{m^{*}\left(
T\right)  }.$

{\small The range of the temperatures for which the theorem holds is a segment
}$\left(  0,\tilde{T}_{c}\left(  d+1\right)  \right)  ,${\small \ except may
be at most countable set of temperature values. It does contain all
temperatures low enough. The value }$\tilde{T}_{c}${\small \ is the so called
``slab percolation threshold'', which is believed to coincide with }$T_{c}.$

For a proof, see \cite{Bo,CeP}, or the review paper \cite{BoIV}.

\subsection{Facets of the Wulff shape and of the random droplet.}

The following is known about the surface tension function $\tau$ (see
(\ref{55})) of a model of statistical mechanics. When the dimension $\left(
d+1\right)  $ of the model is two, then the functions $\tau$ are expected to
be smooth, so they have no cusps, and the corresponding Wulff curves do not
have facets (=straight segments). This is known to be the case for the 2D
Ising model (from the exact solution). The methods of the book \cite{DKS}
allow one to prove that for a general class of the 2D models (so called
Pirogov-Sinai models) the surface tension function $\tau$ is in fact analytic.
The cusps do exist in the low-temperature 3D Ising model surface tension, so
there the Wulff shape indeed has corresponding facets and thus resembles a
crystal. This was proven in \cite{BMF}, and extended later to other models in
\cite{Miracle}. Moreover, it is expected that in the 3D Ising model the so
called roughening transition happens at the temperature $T_{R}<T_{c}$, at
which temperature the surface tension starts to be smooth, the facets
disappear, so the Wulff shape becomes round. This, however, remains a
challenging open problem.

The Theorem \ref{w3} above provides some information as to how closely the
random droplet, forced into the system by a constraint (\ref{60}), resembles
the Wulff shape. However, this result is not strong enough to permit one to
understand whether the big droplet (which is unique in a typical
configuration) itself has facets, in some sense. Using the notation of the
Theorem \ref{w3}, we can formulate the following hypothesis:

\begin{conjecture}
Let the temperature $T\ $be low enough. Then the following event has
$\mu_{\Lambda(l),per,T,0}^{<\rho}$-probability approaching $1$ as
$l\rightarrow\infty:$

\noindent There exist six distinct 2D planes $L_{i}=L_{i}\left(
\sigma\right)  \subset\mathbb{T}^{3},$ $i=1,...,6,$ two for each coordinate
direction, such that the intersections $L_{i}\cap\Gamma\left(  \sigma\right)
$ are flat facets of $\Gamma\left(  \sigma\right)  .$ Namely, for every $i$

\noindent$i)$ $\,\mathrm{diam}\left(  L_{i}\cap\Gamma\left(  \sigma\right)
\right)  \geq C_{1}\left(  T\right)  \mathrm{diam}\left(  \lambda K_{\tau}%
^{<}\right)  l\,$ (see (\ref{61})), with $C_{1}\left(  T\right)
\rightarrow\sqrt{2/3}$ as $T\rightarrow0;$

\noindent$ii)$ $\,\frac{\left|  L_{i}\cap\Gamma\left(  \sigma\right)  \right|
}{\left[  \mathrm{diam}\left(  L_{i}\cap\Gamma\left(  \sigma\right)  \right)
\right]  ^{2}}\geq C_{2}\left(  T\right)  ,$ with $C_{2}\left(  T\right)
\rightarrow1/2$ as $T\rightarrow0.$

\noindent$iii)$ \thinspace The asymptotic shape of the facets $L_{i}\cap
\Gamma\left(  \sigma\right)  $ is given by the corresponding Wulff
construction, described in Sect. 2.5.
\end{conjecture}

{\small As }$T\rightarrow0,${\small \ the Wulff shape tends to a cube. This is
where the value }$\sqrt{2/3}${\small \ in }$i)${\small \ and }$1/2$ in $ii)$
{\small come from.}

\subsection{Metastability and critical droplets}

In this section we describe rigorous results concerning the problem of
convergence to equilibrium. Intuitively one should think about the situation
when one observes liquid water at negative temperature -- i.e. the supercooled
water -- and one is interested to know how the process evolves and how long
does it take for water to freeze.

Our playground will be again the Ising model in two dimensions. Now we need to
supply it with the time evolution. For this we will use what is known as
Glauber dynamics. It is a Markov process on the state space $\Omega$, whose
generator, $L$, acts on a generic local observable $f$ as
\[
(Lf)(\sigma)=\sum_{x\in\mathbb{Z}^{2}}c(x,\sigma)(f(\sigma^{x})-f(\sigma)),
\]
where $\sigma^{x}$ is the configuration obtained from $\sigma$ by flipping the
spin at the site $x$ to the opposite value, and $c(x,\sigma)$ is the rate of
the flip of the spin at the site $x$ when the system is in the state $\sigma$.
In words, one can say that the dynamics proceeds as follows: at every site $x$
the spin $\sigma\left(  x\right)  $ is flipped randomly, independently of all
other, with the rate $c(x,\sigma),$ where $\sigma$ is the current
configuration. Common examples are \textit{Metropolis Dynamics: }
\[
c_{h}(x,\sigma)=\exp(-\beta(\Delta_{x}H_{h}(\sigma))^{+}),
\]
or \textit{Heat Bath Dynamics: }
\[
c_{h}(x,\sigma)=\left[  1+\exp(\beta\Delta_{x}H_{h}(\sigma))\right]  ^{-1}.
\]
Here $(a)^{+}=\max\{a,0\},$ and $\Delta_{x}H_{h}(\sigma)=$ $H_{h}(\sigma
^{x})-H_{h}(\sigma).$ The spin flip system thus obtained will be denoted by
$(\sigma_{T,h;t}^{\xi})_{t\geq0}$, where $\xi$ is the initial configuration at
time $t=0$. If this initial configuration is selected at random according to a
probability measure $\nu$, then the resulting process is denoted by
$(\sigma_{T,h;t}^{\nu})_{t\geq0}$. It is known that the Gibbs measures are
invariant with respect to the stochastic Ising models. Moreover,
\[
\sigma_{T,h;t}^{-}\rightarrow\mu_{-,T,h},\,\,\sigma_{T,h;t}^{+}\rightarrow
\mu_{+,T,h},\,\,\text{as }t\rightarrow\infty.
\]
We will be interested in the case when $h$ is positive, though small. Then
there is only one invariant state, $\mu_{+,T,h},$ so the state $\mu_{-,T,h} $
is equal to $\mu_{+,T,h},$ and $\sigma_{T,h;t}^{-}\rightarrow\mu_{+,T,h},$
$\,$as $t\rightarrow\infty.$ (One should intuitively think about the state
$\sigma_{T,h;t}^{-}$ for $t$ small as the supercooled but liquid water, and
identify the state $\mu_{+,T,h}$ with ice.) We want to control the convergence
of the temporal state $\sigma_{T,h;t}^{-}$ to the equilibrium, $\mu_{+,T,h},$
and to see, if possible, that during some quite long time the state
$\sigma_{T,h;t}^{-}$ looks very similar to the $\left(  -\right)  $-phase
$\mu_{-,T},$ while after some time threshold it changes suddenly and looks
quite similar to the state $\mu_{+,T,h}.$ It turns out that all the above
features can indeed be established rigorously.

If one starts to simulate the above dynamics on a computer, then the picture
observed would be the following: one would see that droplets of the $\left(
+\right)  $-phase are created in the midst of minuses, which droplets are
there for a while, and then disappear. That process goes on for a while, until
a $\left(  +\right)  $-droplet big enough is born. This one then starts to
grow and eventually fills up all the display. One guesses rightly that the
relevant object here is the droplet energy functional $\Phi,$ introduced in
(\ref{41}). We need a two-dimensional version of it, so the set $D$ consists
of all closed curves $\frak{m}$ in $\mathbb{R}^{2}.$ A trivial adjustment is
needed in the volume (=area) term:
\begin{equation}
\Phi\left(  \frak{m}\right)  =\mathcal{W}_{\tau}\left(  \frak{m}\right)
-m^{*}(T)h\,\mathrm{vol}\left(  \frak{m}\right)  .\label{71}%
\end{equation}
It is due to the fact that because of the way the magnetic field $h$ enters in
the Hamiltonian (\ref{70}), the gain in the energy per unit volume obtained by
replacing the $\left(  -\right)  $-phase by the $\left(  +\right)  $-phase is
given by $m^{*}(T)h\,.$ We remind the reader that the functional $\Phi$ has
unique saddle point, which is the curve $\frak{m}_{\mathrm{sdl}}%
=\frak{m}_{\mathrm{sdl}}\left(  T,h\right)  =\frac1{hm^{*}(T)}W_{\tau},$ with
$\Phi\left(  \frak{m}_{\mathrm{sdl}}\right)  =\frac1{2hm^{*}(T)}%
\mathcal{W}_{\tau}\left(  \,W_{\tau}\right)  ,$ see (\ref{72}), (\ref{73}).

Let us define the \textbf{critical time exponent }$\lambda_{c}=\lambda_{c}(T)$
by
\begin{equation}
\frac{\lambda_{c}}h=\frac13\frac{\Phi\left(  \frak{m}_{\mathrm{sdl}}\right)
}T.\label{74}%
\end{equation}

\begin{theorem}
Suppose that $T<T_{c},\,h>0$. Let $\nu$ be either the $\left(  -\right)
$-phase $\mu_{-,T}$ or $\delta_{\left\{  \sigma=-\right\}  }.$ (In fact, any
$\nu$ ``between'' these two states would go.) Then the following happens.

\ \noindent$i)$ If $0<\lambda<\lambda_{c}$, then for each $n\in\{1,2,...\}$
and for each local observable $f$,
\begin{equation}
\mathbb{E}\left(  f\left(  \sigma_{T,h;t=\exp\left\{  \lambda/h\right\}
}^{\nu}\right)  \right)  =\sum_{j=0}^{n-1}b_{j}\left(  f\right)
h^{j}+O\left(  h^{n}\right)  ,\label{z12}%
\end{equation}
where
\[
b_{j}\left(  f\right)  =\lim_{h\rightarrow0-}\frac{d^{j}\langle f\rangle
_{-,T,h}}{dh^{j}}.
\]
(We stress that in the last relation we are using the Gibbs fields
corresponding to the \textbf{negative} values of the magnetic field.) In
particular,
\begin{equation}
\mathbb{E}\left(  \sigma_{T,h;t=\exp\left\{  \lambda/h\right\}  }^{\nu}\left(
0\right)  \right)  =-m^{\ast}(T)+O\left(  h\right)  .\label{z16}%
\end{equation}

\ \noindent$ii)$ If $\lambda>\lambda_{c}$, then for any finite positive $C $
there is a finite positive $C_{1}$ such that for every local observable $f$,
\begin{equation}
\left|  \mathbb{E}\left(  f\left(  \sigma_{T,h;t=\exp\left\{  \lambda
/h\right\}  }^{\nu}\right)  \right)  -\langle f\rangle_{T,h}\right|  \leq
C_{1}\left|  \left|  f\right|  \right|  \exp\left\{  -\frac{C}{h}\right\}
.\label{z17}%
\end{equation}
\end{theorem}

The relation (\ref{z12}) implies that the family of non-equilibrium states
$\left\langle \cdot\right\rangle _{T,h;\lambda}^{\nu},\,h>0,$ defined for
every local observable $f$ by
\[
\left\langle f\right\rangle _{T,h;\lambda}^{\nu}=\mathbb{E}\left(  f\left(
\sigma_{T,h;t=\exp\left\{  \lambda/h\right\}  }^{\nu}\right)  \right)  ,
\]
is a $\mathcal{C}^{\infty}$-continuation of the curve $\left\{  \langle
\cdot\rangle_{-,T,h},\,h\le0\right\}  $ of equilibrium states. This is true
for every $0<\lambda<\lambda_{c}$ and every $\nu\ $as above$.$ The states
$\left\langle \cdot\right\rangle _{T,h;\lambda}^{\nu}$ are
the\textit{\ metastable states} we are looking for. The relations (\ref{z12}),
(\ref{z16}) should be interpreted in the sense that before the time
$\exp\left\{  \lambda_{c}/h\right\}  $ our temporal state is still ``liquid'',
while (\ref{z17}) means that after the time $\exp\left\{  \lambda
_{c}/h\right\}  $ freezing happens.

This theorem was obtained in \cite{SS}. Let us explain the heuristics behind
it. It has two ingredients. The first one is that the transition to the
equilibrium is going via creation of droplets of the $(+)$-phase. The second
one is that once such a droplet is created by a thermal fluctuation, with the
size exceeding a certain critical value, it does not die out, but grows
further, with a speed $v$ of the order of $h.$ (This second belief can be
expected to be correct only in dimension 2.) Let us see how these two
hypothesis can give us the right answer. As we already said, to get to the
equilibrium we have to overcome the energy barrier, given by the functional
$\Phi.$ Subcritical droplets are constantly created by thermal fluctuations in
the metastable phase, but they tend to shrink, as is dictated by the energy
landscape $\Phi$. On the other hand, once a supercritical droplet is created
due to a larger fluctuation, it will grow and drive the system to the stable
phase. Since the minimal height of the barrier is $\Phi\left(  \frak{m}%
_{\mathrm{sdl}}\right)  ,$ one predicts the rate of creation of a critical
droplet with center at a given place to be $\exp\left\{  -\frac{\Phi\left(
\frak{m}_{\mathrm{sdl}}\right)  }T\right\}  .$

Comparing with (\ref{74}), we see that we miss the correct answer,
$\exp\left\{  -\frac13\frac{\Phi\left(  \frak{m}_{\mathrm{sdl}}\right)
}T\right\}  ,$ by a factor of $1/3$. The reason for that is the following.
Note that we are concerned with an infinite system, and we are observing it
through a local function $f$, which depends on the spins in a finite set
$\mathrm{supp}\left(  f\right)  $. For us the system will have relaxed to
equilibrium once $\mathrm{supp}\left(  f\right)  $ is covered by a big droplet
of the $\left(  +\right)  $-phase, which appeared spontaneously somewhere and
then grew, as discussed above. We want to estimate how long we have to wait
for the probability of such an event to be close to one. If we suppose that
the radius of the supercritical droplet grows with a speed $v$, then we can
see that the region in space-time, where a droplet which covers $\mathrm{supp}%
\left(  f\right)  $ at time $t$ could have appeared, is, roughly speaking, a
cone with vertex in $\mathrm{supp}\left(  f\right)  $ and which has as base
the set of points which have time-coordinate $0$ and are at most at distance
$tv$ from $\mathrm{supp}\left(  f\right)  $. The volume of such a cone is of
the order of $(vt)^{2}t$. The order of magnitude of the relaxation time,
$t_{rel}$, before which the region $\mathrm{supp}\left(  f\right)  $ is
unlikely to have been covered by a large droplet and after which the region
$\mathrm{supp}\left(  f\right)  $ is likely to have been covered by it can now
be obtained by solving the equation
\[
(vt_{rel})^{2}\,t_{rel}\,\exp\left\{  -\frac{\Phi\left(  \frak{m}%
_{\mathrm{sdl}}\right)  }T\right\}  \sim1.
\]
This gives us almost what we want:
\begin{equation}
t_{rel}\sim v^{-2/3}\ \exp\left\{  \frac13\frac{\Phi\left(  \frak{m}%
_{\mathrm{sdl}}\right)  }T\right\}  .\label{z14}%
\end{equation}
If we would know that $v\sim Ch,$ then it further gives us that
\begin{equation}
t_{rel}\sim\exp\left\{  \frac13\frac{\Phi\left(  \frak{m}_{\mathrm{sdl}%
}\right)  }T\right\}  .\label{z15}%
\end{equation}
The hypothesis that $v\sim Ch$ seems natural in our context, yet it remains an
interesting open question. Happily, it is clearly seen from the relation
(\ref{z14}), that to conclude (\ref{z15}) from it we need much less. What we
are able to establish in \cite{SS}, and what is enough for our purposes, is
the following weaker estimate: for every $\varepsilon>0$ there is a $C_{1}>0,$
such that
\[
v\ge C_{1}\exp\left\{  -\frac\varepsilon h\right\}  .
\]

\section{Combinatorics}

In this section we consider applications of results of Section 2 to
combinatorics. We discuss the typical shape of a big Young diagram and that of
a random skyscraper (=plane partition). We make a detour in subsection 4.4 and
discuss there the famous MacMahon formula and the $q$-analog of the hook formula.

\subsection{Typical Young diagrams}

A partition $p$ of an integer $N$ is an array of non-negative integers
$n_{1}\ge n_{2}\ge...\ge n_{k}\ge...,$ such that $\sum_{i=1}^{\infty}n_{i}=N.$
It can be specified by the sequence $\left\{  r_{k}\right\}  $ of integers,
with $r_{k}=l$ iff exactly $l$ elements of $p$ equal $k.$ It can also be
described by the monotone function
\begin{equation}
\phi_{p}\left(  y\right)  =\sum_{k=\left\lceil y\right\rceil }^{\infty}%
r_{k},\;y>0.\label{u11}%
\end{equation}
Its graph $G\left[  \phi_{p}\right]  $ provides a graphical description of $p$
and is called a \textit{Young diagram}. The same name \textit{Young diagram}
will be used also for the set $Y=\left\{  \left(  y_{1},y_{2}\right)
:\phi_{p}\left(  y_{1}\right)  >0,\,0<y_{2}<\phi_{p}\left(  y_{1}\right)
\right\}  \subset\mathbb{R}^{2}.$ The number $N$ of the unit cells in $Y$ will
be called the \textit{volume }$\mathrm{vol}\left(  Y\right)  $ of $Y.$ The set
of all possible Young diagrams will be denoted by $\mathcal{Y}.$ The subset of
Young diagrams with $N$ cells will be denoted by $\mathcal{Y}_{N}.$

Similarly, a plane partition $S$ of an integer $N$ is a two-dimensional array
of non-negative integers $n_{ij},$ such that for any $i$ we have $n_{i1}\ge
n_{i2}\ge...\ge n_{ik}\ge...,$ for any $j$ we have $n_{1j}\ge n_{2j}\ge...\ge
n_{kj}\ge...,$ while again $\sum_{i,j=1}^{\infty}n_{ij}=N.$ One defines the
corresponding function $\phi_{S}\left(  y_{1},y_{2}\right)  ,\;y_{1},y_{2}>0,$
in the obvious way. The function $\phi_{S}\left(  y_{1},y_{2}\right)  $ is
monotone in each variable. Its graph $G\left[  \phi_{S}\right]  $ will be
called a \textit{skyscraper}. Sometime we will call by a skyscraper the
partition $S$ itself. Again, $N$ will be called the \textit{volume
}$\mathrm{vol}\left(  S\right)  $ of $S.$ A support, $\mathrm{supp}\left(
S\right)  ,$ is just the support of the function $\phi_{S}.$ The set of all
(finite) skyscrapers will be denoted by $\mathcal{S}.$ If $B\subset
\mathbb{R}_{+}^{2}=\left\{  \left(  y_{1},y_{2}\right)  :y_{1},y_{2}%
>0\right\}  $ is a subset, finite or infinite, then by $\mathcal{S}^{B}%
\subset\mathcal{S}$ we denote the subset of all skyscrapers with support in $B.$

Higher dimensional analogs of the above objects are straightforward. For
example, we can talk about spacial partitions (4D diagrams), etc.

Many more objects of a similar type can be defined. For example, one can put
restrictions on how the steps of the stair $G\left[  \phi_{p}\right]  $ can
look: they can not be longer than 3 units, and their heights can be only 1,2
or 5, say. The same freedom is allowed in 3D, and above.

Let us fix the number $N,$ choose the dimension and the kind of diagrams we
are interested in, and consider the corresponding set $\mathcal{D}_{N}$ of all
these diagrams. There are finitely many of them, so we can put a uniform
probability distribution on $\mathcal{D}_{N}.$ (Here, again, variations are
possible.) The question now is the following: how the typical diagram from the
family $\mathcal{D}_{N}$ looks like, when $N\rightarrow\infty?$

The first problem of that type was solved in the paper \cite{VKer}, see also
\cite{V1,V2, DVZ}. It was found there, that the typical 2D Young diagram under
statistics described above, if scaled by the factor $\left(  1/\sqrt
{N}\right)  ,$ tends to the curve $\mathcal{C}_{VK},$ given by
\begin{equation}
\exp\left\{  -\tfrac{\pi}{\sqrt{6}}x\right\}  +\exp\left\{  -\tfrac{\pi}%
{\sqrt{6}}y\right\}  =1.\label{08}%
\end{equation}
More precisely, for every $\varepsilon>0$ the probability that the scaled
Young diagram would be within distance $\varepsilon$ from the curve (\ref{08})
goes to $1$ as $N\rightarrow\infty.$

The heuristic way to obtain (\ref{08}) (and similar results) is the following:

$i)$ Let $A=\left(  a_{1},a_{2}\right)  ,B=\left(  b_{1},b_{2}\right)  $ be
two points in $\mathbb{Z}^{2}$, with $a_{1}<b_{1},a_{2}>b_{2}.$ We can easily
see that the number $\#\left(  A,B\right)  $ of lattice staircases, starting
from $A$, terminating at $B,$ and allowed to go only to the right or down, is
given by $\binom{\left(  b_{1}-a_{1}\right)  +\left(  a_{2}-b_{2}\right)
}{\left(  b_{1}-a_{1}\right)  }.$ Therefore one concludes by using the
Stirling formula that
\begin{equation}
\lim_{\left|  B-A\right|  \rightarrow\infty}\frac1{\left|  B-A\right|  }%
\ln\#\left(  A,B\right)  =h\left(  \mathbf{n}_{AB}\right)  .\label{24}%
\end{equation}
Here $\mathbf{n}_{AB}$ is the unit vector, normal to the segment $\left[
A,B\right]  ,$ the limit $A,B\rightarrow\infty$ is taken in such a way that
the vector $\mathbf{n}_{AB}$ does not change, and for $\mathbf{n}=\left(
n_{1},n_{2}\right)  ,$ $\alpha=\frac{n_{1}}{n_{1}+n_{2}},$ the \textit{entropy
density function}
\begin{equation}
h\left(  \mathbf{n}\right)  =-\left(  n_{1}\ln\frac{n_{1}}{n_{1}+n_{2}}%
+n_{2}\ln\frac{n_{2}}{n_{1}+n_{2}}\right)  .\label{76}%
\end{equation}

$ii)$ One argues that the number of Young diagrams of the area $N$ scaled by
$\sqrt{N},$ ``going along'' the monotone curve $y=c\left(  x\right)  \ge0$
with integral one, is approximately given by
\begin{equation}
\exp\left\{  \sqrt{N}\int_{0}^{\infty}h\left(  -\tfrac{c^{\prime}\left(
x\right)  }{\sqrt{1+\left(  c^{\prime}\left(  x\right)  \right)  ^{2}}}%
,\tfrac1{\sqrt{1+\left(  c^{\prime}\left(  x\right)  \right)  ^{2}}}\right)
\sqrt{1+\left(  c^{\prime}\left(  x\right)  \right)  ^{2}}dx\right\}
.\label{17}%
\end{equation}

$iii)$ Assuming that indeed the model under consideration exhibits under a
proper scaling some typical behavior, described by a nice smooth non-random
curve (or surface) $\mathcal{C},$ one comes to the conclusion that the curve
$\mathcal{C}$ should be such that the integral in (\ref{17}), computed along
$\mathcal{C},$ is maximal compared with all other allowed curves.

In general case one is not able to write down the corresponding entropy
function precisely. The only information available generally is the existence
of the limit of the type of (\ref{24}), by a subadditivity argument. It should
be stressed that even when the variational problem for the model is known, the
main difficulty of the rigorous treatment of the model is the proof that
indeed it does exhibit a nontrivial behavior after a proper scaling.

The above program was realized in \cite{V1,V2}, see also \cite{DVZ}, for the
2D case described above and for some other cases. In \cite{Bl} a class of more
general 2D problems was studied. The first 3D problem was successfully studied
in \cite{CKeP}. The method of the last paper can also solve the skyscraper
problem, see \cite{Ke}.

One implication of these results is that the asymptotic shapes of various
random combinatorial objects introduced above are given by the construction
described in subsection \ref{2.4} of the present paper, for corresponding
$\eta$ function. For example, in the case of Young diagrams we thus have

\begin{corollary}
In the notation of the Theorem \ref{dw}, the curve $\mathcal{C}_{VK},$ given
by the formula (\ref{08}) coincides with the curve $V_{\eta},$ which results
from the application of our construction (\ref{15}), (\ref{16}), (\ref{18}) to
the function $\eta\left(  \mathbf{n}\right)  =h\left(  \mathbf{n}\right)  $
given by the formulas (\ref{24}), (\ref{76}).
\end{corollary}

Of course, this statement can also be easily checked directly.

Professor Elliot Lieb had suggested to me during our discussion in May, 2000,
that using the relation (\ref{17}) one can recover some results of
Hardy-Ramanujan-Rademacher about the behavior of the function $\left|
\mathcal{Y}_{N}\right|  ,$ i.e. the number of partitions of the integer $N.$
Namely, in the main order in $N$ it has to be
\[
\exp\left\{  \sqrt{N}\mathcal{V}_{h}\left(  \mathcal{C}_{VK}\right)  \right\}
,
\]
where $h$ is given by (\ref{24}), (\ref{76}), and the functional
$\mathcal{V}_{h}$ is defined by (\ref{03}). The integral $\mathcal{V}%
_{h}\left(  \mathcal{C}_{VK}\right)  $ is in fact easy to compute, by using
our Theorem \ref{dw}. Indeed, using its notation, we know from it that
\begin{equation}
\mathcal{C}_{VK}=\sqrt{\frac2{\mathcal{V}_{h}\left(  G_{h}\right)  }}%
G_{h}.\label{81}%
\end{equation}
Therefore
\begin{equation}
\mathcal{V}_{h}\left(  \mathcal{C}_{VK}\right)  =\sqrt{2\mathcal{V}_{h}\left(
G_{h}\right)  }.\label{82}%
\end{equation}
On the other hand, a straightforward computation shows that the curve
$\mathcal{C}_{VK}$ intersects the line $\left\{  x=y\right\}  $ at the point
$\left(  \frac{\sqrt{6}\ln\,2}\pi,\frac{\sqrt{6}\ln\,2}\pi\right)  ,$ while
the curve $G_{h}$ intersects the same line at the point $\frac1{\sqrt{2}%
}\left(  h\left(  \frac1{\sqrt{2}},\frac1{\sqrt{2}}\right)  ,h\left(
\frac1{\sqrt{2}},\frac1{\sqrt{2}}\right)  \right)  =\left(  \ln\,2,\ln
\,2\right)  .$ Hence,
\begin{equation}
\mathcal{C}_{VK}=\frac{\sqrt{6}}\pi G_{h}.\label{80}%
\end{equation}
Comparing (\ref{80}) and (\ref{81}), we find that $\sqrt{2/\mathcal{V}%
_{h}\left(  G_{h}\right)  }=\sqrt{6}/\pi,$ so in view of (\ref{82})%

\[
\mathcal{V}_{h}\left(  \mathcal{C}_{VK}\right)  =\pi\sqrt{\frac23},
\]
to be compared with Hardy-Ramanujan-Rademacher result that
\[
\left|  \mathcal{Y}_{N}\right|  \sim\frac1{4N\sqrt{3}}\exp\left\{  \pi
\sqrt{\frac{2N}3}\right\}  ,
\]
see, for example, \cite{An}.

\subsection{Facets of skyscrapers}

In this section we will discuss the problem of the typical shape of a random
skyscraper, as described in the previous section. As was explained above, this
shape $\frak{S}$ is a (non-random) concave surface in $\mathbb{R}_{+}^{3},$
given by the dual Wulff construction of section \ref{2.4}. (See \cite{Ke} and
especially \cite{CKeP} for more details about this and related asymptotic
shapes.) We will explain now the proof of the following:

\begin{proposition}
The surface $\frak{S}$ has three identical flat pieces (facets); they belong
to three quadrants $\left(  \mathbb{R}_{+}^{2}\right)  _{i},$ $i=1,2,3,$ which
together form the boundary $\partial\mathbb{R}_{+}^{3}.$ Moreover, the flat
curves $\partial\left(  \frak{S}\cap\left(  \mathbb{R}_{+}^{2}\right)
_{i}\right)  \subset\mathbb{R}^{2}$ are just the Vershik-Kerov curves
$\mathcal{C}_{VK}$ (\ref{08}), properly scaled.
\end{proposition}

Indeed, let $\eta\left(  \mathbf{\cdot}\right)  $ be the corresponding entropy
function, defined on $\Delta^{2}=$ $S^{2}\cap\mathbb{R}_{+}^{3}.\ $Its
definition is the following (compare with (\ref{24})): \smallskip

For $\mathbf{n}\in\Delta^{2}$ consider the plane $L_{\mathbf{n}}=\left\{
\mathbf{x}\in\mathbb{R}^{3},\left(  \mathbf{x,n}\right)  =0\right\}  .$ Let
$\Sigma_{\mathbf{n}}$ be the family of all surfaces $Q$ in $\mathbb{R}^{3}$
with the properties:

$i)$ every $Q$ is made from the unit closed 2D plaquettes of the lattice
$\mathbb{Z}^{3},$

$ii)$ the orthogonal projection of $Q$ onto $L_{\mathbf{n}}$ is a homeomorphism.

\noindent Denote by $Q_{\mathbf{n}}$ (one of) the surface(s) from
$\Sigma_{\mathbf{n}}$ which minimizes the distance $\mathrm{dist}\left(
Q,L_{\mathbf{n}}\right)  .$ Introducing now the quantity $\#\left(
B_{R},Q_{\mathbf{n}}\right)  $ as the number of surfaces in $\Sigma
_{\mathbf{n}},$ which coincide with $Q_{\mathbf{n}}$ outside the ball $B_{R}$
of radius $R$ centered at the origin, we define $\eta\left(  \mathbf{\cdot
}\right)  $ by
\[
\eta\left(  \mathbf{n}\right)  =\lim_{R\rightarrow\infty}\frac{\ln\left(
\#\left(  B_{R},Q_{\mathbf{n}}\right)  \right)  }{\pi R^{2}}.
\]
We will argue now that the derivative $\eta^{\prime}\left(  \mathbf{\nu
}\right)  $ of the function $\eta\left(  \mathbf{\cdot}\right)  $ at the point
$\mathbf{e}_{z}\mathbf{=}\left(  0,0,1\right)  $ in the direction
$\mathbf{\nu}\in\mathbb{R}_{+}^{2},$ tangential to $S^{2}\subset\mathbb{R}%
^{3}$, is nothing else but $h\left(  \mathbf{\nu}\right)  ,$ see (\ref{24}).
Once this is established, the claim about the shape of the facets of the
surface $\frak{S} $ follows by applying our dual Wulff construction, see
(\ref{83}).

To see that $\eta^{\prime}\left(  \mathbf{\nu}\right)  =h\left(  \mathbf{\nu
}\right)  ,$ let $\mathbf{n}_{\theta}\in\Delta^{2}$ be the vector defined by
two properties: $\mathbf{n}_{\theta}=a\mathbf{e}_{z}+b\mathbf{\nu}$ for some
$a,b$ real; and $\left(  \mathbf{n}_{\theta},\mathbf{e}_{z}\right)
=\cos\theta.$ In words, the vector $\mathbf{n}_{\theta}$ is a result of going
from $\mathbf{e}_{z}$ in the direction $\mathbf{\nu}$ by a distance $\theta.$
Since $\eta\left(  \mathbf{e}_{z}\right)  =0,$ we have to consider the limit
$\lim_{\theta\rightarrow0}\frac{\eta\left(  \mathbf{n}_{\theta}\right)
}\theta\equiv\eta^{\prime}\left(  \mathbf{\nu}\right)  .$

Let $\theta$ be small. Then the surface $Q_{\mathbf{n}},$ viewed from the $z $
axis, can be represented by the collection of periodic lattice staircases
$...,s_{-1},s_{0},s_{1},...$ , with the spacing between two consecutive $s$-s
of the order of $\left(  \sin\theta\right)  ^{-1}.$ If a surface $Q$
contributes to $\#\left(  B_{R},Q_{\mathbf{n}}\right)  ,$ then it is also
represented by the sequence of lattice staircases $...,s_{-1}^{\prime}%
,s_{0}^{\prime},s_{1}^{\prime},...$ , with the following properties:

\noindent$i)$ only finitely many of $s_{i}^{\prime}$ are different from the
corresponding $s_{i};$

\noindent$ii)$ each $s_{i}^{\prime}$ coincides with its correspondent, $s_{i},
$ outside $B_{R}$;

\noindent$iii)$ different $s_{i}^{\prime}$-s might touch each other and have
common bonds, but they cannot cross each other.

\noindent The removal of the third restriction makes the stairs independent,
and that immediately provides us with the upper bound
\[
\eta^{\prime}\left(  \mathbf{\nu}\right)  \le h\left(  \mathbf{\nu}\right)  .
\]
To obtain the opposite inequality, let us introduce the modified functions
$h_{\varepsilon}\left(  \mathbf{\cdot}\right)  ,$ which are defined similarly
to the function $h\left(  \mathbf{\cdot}\right)  $ (see (\ref{24})), with the
extra restriction that only those stairs are allowed which deviate from the
segment $\left[  A,B\right]  $ by at most $\varepsilon^{-1}/2.$ Consider now a
collection $...,s_{-1}^{\prime},s_{0}^{\prime},s_{1}^{\prime},...$ of lattice
staircases, which satisfy the properties $i),$ $ii)$ from above and the
following property:

\noindent$iii^{\prime})$ each stair $s_{i}^{\prime}$ can deviate away from its
correspondent, $s_{i},$ by at most $\left(  \sin\theta\right)  ^{-1}/2.$

\noindent Clearly, every such collection defines a surface, $Q\left(
...,s_{-1}^{\prime},s_{0}^{\prime},s_{1}^{\prime},...\right)  ,$ which
contributes to $\#\left(  B_{R},Q_{\mathbf{n}}\right)  .$ That immediately
implies that for every $\theta>0$%
\[
h_{\sin\theta}\left(  \mathbf{\nu}\right)  \le\eta^{\prime}\left(
\mathbf{\nu}\right)  .
\]
Since $h_{\sin\theta}\left(  \mathbf{\nu}\right)  \nearrow h\left(
\mathbf{\nu}\right)  $ as $\theta\rightarrow0$ (see, for example,
\cite{Messager}), that proves the equality $\eta^{\prime}\left(  \mathbf{\nu
}\right)  =h\left(  \mathbf{\nu}\right)  .$

\subsection{Independent variables representation and some generating functions
\label{4.3}}

In this section we will describe one technical device which can be used to
prove some of the above combinatorial statements.

We will call a (plane) partition \textit{strict} iff all the entries of the
(two-dimensional) array are distinct. In what follows, we will denote by
$\pi_{2}\left(  N\right)  $ ($\pi_{2}^{s}\left(  N\right)  $) the number of
different (strict) partitions of the integer $N,$ and by $\pi_{3}\left(
N\right)  $ ($\pi_{3}^{s}\left(  N\right)  $) the number of (strict) plane
partitions of $N.$

It is not difficult to see that the generating function $\frak{g}_{2}\left(
x\right)  $ of the sequence $\left\{  \pi_{2}\left(  i\right)
,i=1,2,...,N,...\right\}  $ is given by
\begin{equation}
\frak{g}_{2}\left(  x\right)  =\prod_{l=1}^{\infty}\frac1{1-x^{l}}.\label{ug2}%
\end{equation}
Indeed, after expanding each factor of (\ref{ug2}), we obtain the following
expression:
\[
\frak{g}_{2}\left(  x\right)  =\sum_{r_{1},r_{2},r_{3},...}x^{r_{1}}x^{2r_{2}%
}x^{3r_{3}}...,
\]
where the summation is taken over all infinite sequences $\left\{  r_{1}%
,r_{2},r_{3},...\right\}  $ of nonnegative integers, which have only finitely
many non-zero entries. Each sequence of $r$-s defines the Young diagram by the
relation (\ref{u11}). In the same way, the generating function of the sequence
$\pi^{s}\left(  N\right)  $ is given by
\[
\frak{g}_{2}^{s}\left(  x\right)  =\prod_{l=1}^{\infty}\left(  1+x^{l}\right)
.
\]

From the point of view of statistical mechanics the expression (\ref{ug2}) is
nothing else but the \textit{grand canonical partition function} of a
one-dimensional system of non-interacting spin particles $\zeta_{l},$
$l=1,2,...,$ with $\zeta_{l}=0,1,2,...$ $.$ The particles $\zeta_{l}$ are not
identical; the statistical weight for the $l$-th particle to be in the state
$k$ is equal to $\left(  x^{l}\right)  ^{k}.$ The parameter $x$ plays here the
role of chemical potential. Let us associate with every particle $\zeta_{l}$ a
volume $v\left(  \zeta_{l}\right)  ,$ so when the particle is in the state $k,
$ its volume will be $lk.$ The statistical weight $w_{x}\left(  \mathbf{\zeta
}\right)  $ of a configuration $\mathbf{\zeta}=\zeta_{1},\zeta_{2,}\zeta
_{3},...$ is now given by
\begin{equation}
w_{x}\left(  \mathbf{\zeta}\right)  =x^{v\left(  \zeta_{1}\right)  +v\left(
\zeta_{2}\right)  +v\left(  \zeta_{3}\right)  +...}.\label{u30}%
\end{equation}
The fact that the particles are not interacting is manifested by the absence
of cross-terms in the exponent in (\ref{u30}). The product $\left(
\frak{g}_{2}\left(  x\right)  \right)  ^{-1}w_{x}\left(  \mathbf{\zeta
}\right)  $ defines a probability distribution on $\mathbf{\zeta}$-s. This is
what is called \textit{grand canonical ensemble} in the language of
statistical mechanics.

Note now that if $x<1,$ then the distribution (\ref{u30}) allows only
configurations $\mathbf{\zeta}$ with finite total volume $\mathrm{vol}\left(
\mathbf{\zeta}\right)  =v\left(  \zeta_{1}\right)  +v\left(  \zeta_{2}\right)
+v\left(  \zeta_{3}\right)  +...$ \thinspace. That means that only finitely
many of the values $\zeta_{1},\zeta_{2},\zeta_{3},...$ can be non-zero. Note
also that from every such configuration $\mathbf{\zeta}$ we can build a Young
diagram $Y\left(  \mathbf{\zeta}\right)  $ in the following way: we first put
$\zeta_{1}$ blocks $1\times1$ in a horizontal array, we then put $\ \zeta_{2}$
blocks $1\times2$ next to them (here $2$ is the height of the block), followed
by $\zeta_{3}$ blocks $1\times3,$ and so on, until we use up all non-zero
entries of $\mathbf{\zeta}$. If we place these blocks from right to left, we
get a genuine Young diagram, $Y\left(  \mathbf{\zeta}\right)  .$ The
(normalized) distribution (\ref{u30}) defines therefore for every $x$ a
probability distribution $\mathbb{P}_{x}$ on $\mathcal{Y}=\left\{  Y\right\}  .$

With every \textit{grand canonical} distribution one can associate a
\textit{canonical} distribution, which in our case is the conditional
distribution $\mathbb{P}_{x}\left(  \cdot\,|\,\mathrm{vol}\left(
\cdot\right)  =N\right)  $ on $\mathcal{Y},$ obtained from $\mathbb{P}_{x}$ by
restricting it to the subset $\mathcal{Y}_{N}=\left\{  Y\subset\mathcal{Y}%
:\mathrm{vol}\left(  Y\right)  =N\right\}  $ and normalizing to probability
distribution. The crucial observation now is the fact that the canonical
measure $\mathbb{P}_{x}\left(  \cdot\,|\,\mathrm{vol}\left(  \cdot\right)
=N\right)  $ does not depend on $x $ and is moreover a uniform distribution on
$\mathcal{Y}_{N}.$ This is evident from (\ref{u30}). Therefore the question
about the behavior of the typical diagram, asked in section 4.1, is the
question concerning the properties of a certain canonical ensemble.

At this moment the use of the jargon of statistical mechanics pays back. The
stat-mechanical folklore immediately recognize our question as standard, and
in order to answer it one should use the standard recipe: the principle of
equivalence of ensembles! This general principle in our case claims, roughly,
that the properties of a typical Young diagram are the same in the ensembles
$\mathbb{P}_{x}\left(  \cdot\,|\,\mathrm{vol}\left(  \cdot\right)  =N\right)
$ and in the ensembles $\mathbb{P}_{x}.$ More precisely, the typical
properties of the Young diagrams under $\mathbb{P}_{x}\left(  \cdot
\,|\,\mathrm{vol}\left(  \cdot\right)  =N\right)  ,$ asymptotically, as
$N\rightarrow\infty,$ are the same as those of the Young diagrams under
$\mathbb{P}_{x\left(  N\right)  },$ where the value $x\left(  N\right)  $
satisfies
\[
\mathbb{E}_{\mathbb{P}_{x\left(  N\right)  }}\left(  \mathrm{vol}\left(
Y\right)  \right)  =N.
\]
In words, the value $x\left(  N\right)  $ has to be such that the mean value
of the random variable $\mathrm{vol}\left(  Y\right)  $ under $\mathbb{P}%
_{x\left(  N\right)  }$ is the same as the (non-random) value of
$\mathrm{vol}\left(  Y\right)  $ under $\mathbb{P}_{x}\left(  \cdot
\,|\,\mathrm{vol}\left(  \cdot\right)  =N\right)  ,$ i.e. $N.$ The advantages
of passing to $\mathbb{P}_{x\left(  N\right)  }$ are evident: the random
variables $\zeta_{l}$ are independent under $\mathbb{P}_{x},$ for every $x.$
Of course, there is price to be paid: one has to prove first that the
equivalence of ensembles indeed holds. But this is a standard, though sometime
hard, work. This is exactly the approach used in \cite{DVZ}. The same can be
done in the case of the strict partitions, see again \cite{DVZ}; one ends up
with collection of non-interacting particles $\zeta_{l},$ which however are
taking now only values $0$ and $1.$ More general cases can be treated by
allowing the particles $\zeta_{l}$ to interact via some local interaction, see
\cite{Bl}.

\subsection{More generating functions and the MacMahon formulas.}

Inspired by what was said in the previous section, one would like to repeat
the same program for the plane partitions. The generating function
$\frak{g}_{3}\left(  x\right)  $ of the sequence $\pi_{3}\left(  N\right)  $
is again known. It is given by the famous MacMahon formula (\ref{u31}) below.
Before discussing it we will make some preparatory comments.

Let $S$ be some plane partition of $N,$ or a skyscraper. Then we can define a
partition
\begin{equation}
p=\Pi\left(  S\right) \label{u12}%
\end{equation}
of $N$ (or a Young diagram), by just listing all the entries of the
two-dimensional array $S$ in the non-increasing order. One would like to study
the inverse map, which to every partition $p$ of $N$ corresponds the preimage
$\Pi^{-1}\left(  p\right)  ,$ in hope to obtain in this way the method of
counting plane partitions. The difficulty here lies in the fact that the
preimages $\Pi^{-1}\left(  p_{1}\right)  ,\Pi^{-1}\left(  p_{2}\right)  $
might contain different number of plane partitions, even when $p_{1},p_{2}$
are partitions of $N$ with the same numbers of summands. The reason is that
the partitions $p_{i}$ may contain repeating integers, and the smaller is
their number the bigger is the preimage, being maximal for strict partitions.

We still want to implement this idea of counting skyscrapers via counting
Young diagrams. So we will introduce a subclass of plane partitions, which we
will call \textit{pedestals}. To define them we first have to fix a complete
ordering $\left(  i_{1},j_{1}\right)  \succ\left(  i_{2},j_{2}\right)
\succ...\succ\left(  i_{k},j_{k}\right)  \succ...$ on the set $\mathbb{Z}%
_{+}^{2}=\left\{  \left(  i,j\right)  :i,j>0\right\}  .$ The choice of the
ordering is not important, once it is red out from some strict plane
partition. More precisely, the ordering $\left(  i_{1},j_{1}\right)
\succ\left(  i_{2},j_{2}\right)  \succ...\succ\left(  i_{k},j_{k}\right)
\succ...$ will be called allowed, if the following is satisfied: for every
integer $k$ the array
\[
n_{\left(  i_{l},j_{l}\right)  }=\left\{
\begin{array}
[c]{cc}%
k-l & \text{ if }k\ge l\\
0 & \text{ otherwice}%
\end{array}
\right.
\]
is a plane partition (of a corresponding integer). In particular, the first
site $\left(  i_{1},j_{1}\right)  $ in every allowed ordering has to be the
site $\left(  1,1\right)  .$ We choose for the ordering $\succ$ any allowed
ordering, and it will be fixed in what follows.

Let now $S$ be some skyscraper. Using the ordering $\succ$, we can define
another one, $\succ_{S},$ in the following way:%

\[
\left(  i^{\prime},j^{\prime}\right)  \succ_{S}\left(  i^{\prime\prime
},j^{\prime\prime}\right)  \Longleftrightarrow\left\{
\begin{array}
[c]{ll}%
n_{i^{\prime}j^{\prime}}>n_{i^{\prime\prime}j^{\prime\prime}}, & \text{ or}\\
n_{i^{\prime}j^{\prime}}=n_{i^{\prime\prime}j^{\prime\prime}}, & \text{ but
}\left(  i^{\prime},j^{\prime}\right)  \succ\left(  i^{\prime\prime}%
,j^{\prime\prime}\right)  .
\end{array}
\right.  \text{ }
\]
Clearly, different skyscrapers $S^{\prime}$ and $S^{\prime\prime}$ can lead to
the same orderings $\left(  \succ_{S^{^{\prime}}}\right)  =\left(
\succ_{S^{\prime\prime}}\right)  .$ In such a case we will call them
equivalent, $S^{\prime}\sim S^{\prime\prime}.$

We denote by $\mathcal{O}$ the set of all orderings, defined by skyscrapers:
$\mathcal{O}=\left\{  \succ_{S},S\in\mathcal{S}\right\}  .$

A skyscraper $\frak{P}$ will be called \textit{a pedestal}, iff every
skyscraper $S\sim\frak{P}$ is higher than $\frak{P}.$ That means that if
$S=\left\{  n_{ij}\right\}  ,$ and $\frak{P}=\left\{  \widetilde{n}%
_{ij}\right\}  ,$ then $n_{ij}\ge\widetilde{n}_{ij}$ for every pair $\left(
i,j\right)  .$ In this case we will write $\frak{P}=\frak{P}\left(  \succ
_{S}\right)  =\frak{P}\left(  S\right)  ,$ and we call $\frak{P}$ a pedestal
of $S.$

A more direct definition of the pedestal, which also proves its existence, is
the following: let $\curlyeqsucc$ be some allowed ordering, which coincides
with $\succ$ at infinity; in other words, $\curlyeqsucc\in\mathcal{O}$. The
pedestal $\frak{P}\left(  \curlyeqsucc\right)  =\left\{  \widetilde{n}%
_{ij}\right\}  $ is defined as follows:

\noindent$i)$ for every two sites $\left(  i^{\prime},j^{\prime}\right)
\curlyeqsucc\left(  i^{\prime\prime},j^{\prime\prime}\right)  \in
\mathbb{Z}_{+}^{2},$ with the pair $\left(  i^{\prime},j^{\prime}\right)  $
directly preceding the pair $\left(  i^{\prime\prime},j^{\prime\prime}\right)
$ under $\curlyeqsucc,$ we have
\[
\widetilde{n}_{i^{\prime}j^{\prime}}=\left\{
\begin{array}
[c]{cc}%
\widetilde{n}_{i^{\prime\prime}j^{\prime\prime}}, & \text{ if }\left(
i^{\prime},j^{\prime}\right)  \succ\left(  i^{\prime\prime},j^{\prime\prime
}\right)  ,\\
\widetilde{n}_{i^{\prime\prime}j^{\prime\prime}}+1, & \text{ if }\left(
i^{\prime},j^{\prime}\right)  \prec\left(  i^{\prime\prime},j^{\prime\prime
}\right)  ;
\end{array}
\right.
\]

\noindent$ii)$ the skyscraper $\frak{P}\left(  \curlyeqsucc\right)  $ has zero
height at infinity.

\noindent In words, the height of $\frak{P}\left(  \curlyeqsucc\right)  $ at
the site $\left(  i,j\right)  $ equals to the number of times the ordering
$\curlyeqsucc$ disagree with the ordering $\succ$ once we proceed from
$\left(  i,j\right)  $ to infinity, following the ordering $\curlyeqsucc.$ (By
definition, this number is finite.)

The set of all possible pedestals $\frak{P}$ will be denoted by $\mathcal{P}.
$ Note that by construction, the mapping $\frak{P}\rightsquigarrow
\succ_{\frak{P}}$ defines a one-to-one correspondence between $\mathcal{P}$
and $\mathcal{O}.$

We are now in the position to define a bijection,
\begin{equation}
b_{St}:\mathcal{S}\rightarrow\mathcal{P\,}\times\,\mathcal{Y}.\label{u13}%
\end{equation}
Namely, to a skyscraper $S$ we correspond a pair, $\left(  \frak{P},Y\right)
, $ where $\frak{P}=\frak{P}\left(  S\right)  $ is the pedestal of $S,$ while
\[
Y=p\left(  S-\frak{P}\left(  S\right)  \right)  ,
\]
see (\ref{u12})$.$ Here we treat $S$ and $\frak{P}\left(  S\right)  $ as
real-valued functions, and $S-\frak{P}\left(  S\right)  $ is just their
difference. By definition, $S-\frak{P}\left(  S\right)  $ is again a
skyscraper, so reading its entries in decreasing order produces a Young
diagram $Y$.

The inverse map, $\left(  b_{St}\right)  ^{-1}:\mathcal{P}\times
\mathcal{Y}\rightarrow\mathcal{S},$ is described as follows. To every ordering
$\curlyeqsucc\in\mathcal{O}$ and every partition $p=\left\{  n_{1}\ge n_{2}%
\ge...\ge n_{k}\ge...\right\}  $ we first assign an intermediate skyscraper
$S\left(  \curlyeqsucc,p\right)  $ - by\textit{\ putting }$p$ \textit{along
}$\curlyeqsucc.$ More precisely, we define $S\left(  \curlyeqsucc,p\right)  $
to be the plane partition, which equals to $n_{k}$ at the site $\left(
i,j\right)  \in\mathbb{Z}_{+}^{2}$ which comes $k$-th according to the
ordering $\curlyeqsucc.$ Now we define
\[
\left(  b_{St}\right)  ^{-1}\left(  \frak{P},p\right)  =\frak{P}+S\left(
\succ_{\frak{P}},p\right)  .
\]
The bijection (\ref{u13}) and its generalization (\ref{u15}) which follows are
in fact very close to Stanley bijection, \cite{St}, which explains our notation.

Note that if $b_{St}\left(  S\right)  =\left(  \frak{P},Y\right)  ,$ then by
construction $\mathrm{vol}\left(  S\right)  =\mathrm{vol}\left(
\frak{P}\right)  +\mathrm{vol}\left(  Y\right)  .$ Therefore the generating
function $\frak{g}_{3}\left(  x\right)  $ of skyscrapers is just a product:%

\begin{equation}
\frak{g}_{3}\left(  x\right)  =\frak{g}_{2}\left(  x\right)  \times
\frak{g}_{\mathcal{P}}\left(  x\right)  .\label{u14}%
\end{equation}
Here $\frak{g}_{\mathcal{P}}\left(  x\right)  =\sum_{k=0}^{\infty}a_{k}x^{k}$
is the generating function of the pedestals: the number $a_{k}$ equals to the
number of pedestals with volume $k.$

Note now, that everything which was said above, about the set $\mathcal{S}$ of
skyscrapers built over $\mathbb{Z}_{+}^{2},$ can be literally repeated for the
subfamily $\mathcal{S}^{B}$ of skyscrapers satisfying the restriction that
their supports belong to a fixed finite subset $B\subset\mathbb{Z}_{+}^{2}.$
(The only thing one has to be careful with is the ordering $\succ:$ one has to
assume that according to it each point in $B$ precedes each point in
$\mathbb{Z}_{+}^{2}\,\backslash\,B.$ That puts a restriction on the shape of
$B:$ it has necessarily to have the shape of a Young diagram.) Their
generating function will be denoted by $\frak{g}_{3}^{B}\left(  x\right)  ,$
the corresponding set of all orderings $\mathcal{O}^{B}$ will be a subset of
$\mathcal{O},$ while the set of corresponding pedestals $\mathcal{P}^{B}$
satisfies $\mathcal{P}^{B}=\mathcal{P}\cap\mathcal{S}^{B}.$ Finally, the
bijection $b_{St}$ is generalized to
\begin{equation}
b_{St}^{B}:\,\mathcal{S}^{B}\rightarrow\,\mathcal{P}^{B}\mathcal{\,}%
\times\,\mathcal{Y}^{\left|  B\right|  },\label{u22}%
\end{equation}
where $\mathcal{Y}^{\left|  B\right|  }$ is the set of all partitions with at
most $\left|  B\right|  $ nonzero entries. If we denote the generating
function of the set of partitions belonging to $\mathcal{Y}^{\left|  B\right|
}$ by $\frak{g}_{2}^{\left|  B\right|  }\left(  x\right)  ,$ then we have
\begin{equation}
\frak{g}_{2}^{\left|  B\right|  }\left(  x\right)  =\prod_{l=1}^{\left|
B\right|  }\frac1{1-x^{l}}.\label{u33}%
\end{equation}
The relation (\ref{u14}) becomes
\begin{equation}
\frak{g}_{3}^{B}\left(  x\right)  =\,\frak{g}_{2}^{\left|  B\right|  }\left(
x\right)  \times\frak{g}_{\mathcal{P}^{B}}\left(  x\right)  .\label{u23}%
\end{equation}
Note that for finite $B$ the set $\mathcal{P}^{B}$ of the pedestals is also
finite, and therefore the function $\frak{g}_{\mathcal{P}^{B}}\left(
x\right)  $ is just a polynomial.

The next generalization we need is to consider skyscrapers $\phi_{S}\left(
y_{1},y_{2}\right)  $ which are not defined everywhere on $\mathbb{Z}_{+}^{2}%
$. That is, we fix a finite subset $A\subset\mathbb{Z}_{+}^{2},$ and we
consider functions $\phi_{S}\left(  y_{1},y_{2}\right)  $ on $\mathbb{Z}%
_{+}^{2}\,\backslash\,A$ satisfying all the properties of the skyscraper
function. All the objects of the previous paragraph can be introduced in this
new context as well. (Again, we want the ordering $\succ\ $to be such that
according to it each point in $A$ precedes each point in $\mathbb{Z}_{+}%
^{2}\,\backslash\,A.$ This forces $A$ to be Young-diagram-shaped.)

Finally, we put both constructions together, considering a pair of Young
diagrams $A\subset B\subset\mathbb{Z}_{+}^{2},$ and the corresponding set
$\mathcal{S}^{AB}$ of skyscrapers defined on $\mathbb{Z}_{+}^{2}%
\,\backslash\,A,$ which vanish outside $B.$ The ordering $\succ\ $should be
such that each point in $A$ precedes each point in $B\,\backslash\,A,$ while
each point in $B$ precedes each point in $\mathbb{Z}_{+}^{2}\,\backslash\,B.$
The generating function for $\mathcal{S}^{AB}$ will be denoted by
$\frak{g}_{3}^{AB}\left(  x\right)  ;$ we have also orderings $\mathcal{O}%
^{AB}$, pedestals $\mathcal{P}^{AB}$ and the bijection
\begin{equation}
b_{St}^{AB}:\,\mathcal{S}^{AB}\rightarrow\,\mathcal{P}^{AB}\mathcal{\,}%
\times\,\mathcal{Y}^{\left|  B\backslash A\right|  }.\label{u15}%
\end{equation}
The relation (\ref{u14}) is further generalized to
\begin{equation}
\frak{g}_{3}^{AB}\left(  x\right)  =\,\frak{g}_{\mathcal{P}^{AB}}\left(
x\right)  \prod_{l=1}^{\left|  B\backslash A\right|  }\frac1{1-x^{l}%
},\label{u16}%
\end{equation}
again with the polynomial generating function $\frak{g}_{\mathcal{P}^{AB}%
}\left(  x\right)  $.

There is another remarkable expression for the function $\frak{g}_{3}%
^{AB}\left(  x\right)  ,$ but \textit{only in case }when\textit{\ }the
diagram\textit{\ }$B$ is a rectangle,\textit{\ }$B=m\times n.$ If
$A=\emptyset,$ then it is the famous MacMahon formula:
\begin{equation}
\frak{g}_{3}^{m\times n}\left(  x\right)  =\prod_{i=1}^{m}\,\prod_{j=1}%
^{n}\frac1{1-x^{i+j-1}},\label{u17}%
\end{equation}
which is also valid for $B=\mathbb{Z}_{+}^{2}:$%
\begin{equation}
\frak{g}_{3}\left(  x\right)  =\prod_{i=1}^{\infty}\,\prod_{j=1}^{\infty}%
\frac1{1-x^{i+j-1}}.\label{u31}%
\end{equation}
For general diagram $A\subset B$ (\ref{u17}) is generalized as follows:
\begin{equation}
\frak{g}_{3}^{AB}\left(  x\right)  =\prod_{b\in B\backslash A}\frac
1{1-x^{h\left(  b\right)  }}.\label{u18}%
\end{equation}
Here the product is taken over all $\left|  B\,\backslash\,A\right|  $ unit
cells of the diagram $B\,\backslash\,A,$ while $h\left(  b\right)  $ is the
hook length of the hook of the cell $b:$ it is the number of cells of the
diagram $B\,\backslash\,A$ in the column of $b$ below $b$ plus the number of
cells in the row of $b$ to the left of $b$ plus $1.$ The relation (\ref{u18})
can be found in \cite{HG}.

Comparing the two expressions, (\ref{u18}) and (\ref{u16}), we see that the
following holds:

\begin{proposition}
The polynomial $\frak{g}_{\mathcal{P}^{AB}}\left(  x\right)  $ satisfies
\begin{equation}
\frak{g}_{\mathcal{P}^{AB}}\left(  x\right)  =\frac{\prod_{l=1}^{\left|
B\backslash A\right|  }\left(  1-x^{l}\right)  }{\prod_{b\in B\backslash
A}\left(  1-x^{h\left(  b\right)  }\right)  }.\label{u19}%
\end{equation}
In particular, for the rectangle we have
\begin{equation}
\frak{g}_{\mathcal{P}^{m\times n}}\left(  x\right)  =\frac{\prod_{l=1}%
^{mn}\left(  1-x^{l}\right)  }{\prod_{i=1}^{m}\,\prod_{j=1}^{n}\left(
1-x^{i+j-1}\right)  }.\label{uu20}%
\end{equation}
\end{proposition}

The fact that in (\ref{u19}) and (\ref{uu20}) we have complete cancellations,
and the rational functions in the r.h.s. are in fact polynomials looks quite surprising!

In the limit as $x\rightarrow1$ we get
\[
\frak{g}_{\mathcal{P}^{AB}}\left(  1\right)  =\frac{\left(  \left|
B\backslash A\right|  \right)  !}{\prod_{b\in B\backslash A}h\left(  b\right)
}.
\]
This is the famous hook formula for the dimension of the irreducible
representation of the symmetric group with $\left|  B\backslash A\right|  $
elements, corresponding to the diagram $B\backslash A.$ The relation
(\ref{u19}) is its $q$-analog. It was proven in \cite{Ker}, \cite{Kr}, though
it was known earlier. Our proof via using (\ref{u16}) and (\ref{u18}) seems to
be the simplest known.

Let us now discuss the higher dimensional situation. Consider the function
$\frak{g}_{4}^{m\times n\times k}\left(  x\right)  ,$ which is the generating
function of the spatial partitions, i.e. 3D arrays with obvious decay
restrictions. Long time ago MacMahon had conjectured, in analogy with
(\ref{u33}), (\ref{uu20}), that
\[
\frak{g}_{4}^{m\times n\times k}\left(  x\right)  =\prod_{i=1}^{m}%
\,\prod_{j=1}^{n}\prod_{l=1}^{k}\frac1{1-x^{i+j+l-2}}.
\]
Later he started to doubt it, and then it turned out that this formula is not
correct, see \cite{An}. However, the higher dimensional analogs of the
bijections (\ref{u13}) (\ref{u15}) as well as the product representation
(\ref{u23}) are always valid. For example, we have for the parallelepipeds
$2\times2\times2$ and $2\times2\times3$ that
\begin{equation}
\frak{g}_{4}^{2\times2\times2}\left(  x\right)  =\frac{\frak{g}_{\mathcal{P}%
^{2\times2\times2}}\left(  x\right)  }{\prod_{l=1}^{8}\left(  1-x^{l}\right)
},\label{u24}%
\end{equation}%
\begin{equation}
\frak{g}_{4}^{2\times2\times3}\left(  x\right)  =\frac{\frak{g}_{\mathcal{P}%
^{2\times2\times3}}\left(  x\right)  }{\prod_{l=1}^{12}\left(  1-x^{l}\right)
},\label{u25}%
\end{equation}
where the polynomials $\frak{g}_{\mathcal{P}^{2\times2\times2}}\left(
x\right)  $ and $\frak{g}_{\mathcal{P}^{2\times2\times3}}\left(  x\right)  $
are:
\begin{align*}
\frak{g}_{\mathcal{P}^{2\times2\times2}}\left(  x\right)   & =1+2x^{2}%
+2x^{3}+3x^{4}+3x^{5}+5x^{6}+4x^{7}+8x^{8}+\\
& 4x^{9}+5x^{10}+3x^{11}+3x^{12}+2x^{13}+2x^{14}+x^{16},
\end{align*}%
\begin{align*}
\frak{g}_{\mathcal{P}^{2\times2\times3}}\left(  x\right)   & =1+2x^{2}%
+3x^{3}+5x^{4}+6x^{5}+12x^{6}+14x^{7}+\\
& +25x^{8}+29x^{9}+41x^{10}+46x^{11}+60x^{12}+68x^{13}+86x^{14}+\\
& +96x^{15}+117x^{16}+123x^{17}+141x^{18}+137x^{19}+144x^{20}+\\
& +140x^{21}+144x^{22}+137x^{23}+141x^{24}+123x^{25}+117x^{26}+\\
& +96x^{27}+86x^{28}+68x^{29}+60x^{30}+46x^{31}+41x^{32}+29x^{33}+\\
& +25x^{34}+14x^{35}+12x^{36}+6x^{37}+5x^{38}+3x^{39}+2x^{40}+x^{42},
\end{align*}
as one can find with bare hands. (In fact, the above computations of the
polynomials $\frak{g}_{\mathcal{P}^{2\times2\times2}}\left(  x\right)  $ and
$\frak{g}_{\mathcal{P}^{2\times2\times3}}\left(  x\right)  $ were done
together with Oleg Ogievecky, using Mathematica.)

One then checks that the polynomial $\prod_{l=1}^{8}\left(  1-x^{l}\right)  $
is not divisible by $\frak{g}_{\mathcal{P}^{2\times2\times2}}\left(  x\right)
, $ nor the polynomial $\prod_{l=1}^{12}\left(  1-x^{l}\right)  $ by
$\frak{g}_{\mathcal{P}^{2\times2\times3}}\left(  x\right)  .$ That shows that,
in some sense, there is no higher-dimensional analogs of the MacMahon formula.
However, by (\ref{u24}), (\ref{u25}) we can still write down the relevant
generating functions, having for example that%

\begin{align*}
\frak{g}_{4}^{2\times2\times2}\left(  x\right)   & =1+x+4x^{2}+7x^{3}%
+14x^{4}+23x^{5}+\\
& 41x^{6}+63x^{7}+104x^{8}+152x^{9}+230x^{10}+...,
\end{align*}%
\begin{align*}
\frak{g}_{4}^{2\times2\times3}\left(  x\right)   & =1+x+4x^{2}+8x^{3}%
+17x^{4}+30x^{5}+58x^{6}+97x^{7}+\\
& +171x^{8}+276x^{9}+450x^{10}+...\;.
\end{align*}

\subsection{Independent variables representation for the skyscrapers.}

Now we can, using MacMahon formula
\[
\frak{g}_{3}\left(  x\right)  =\prod_{i=1}^{\infty}\,\prod_{j=1}^{\infty}%
\frac1{1-x^{i+j-1}},
\]
try to obtain the independent variables representation for the skyscrapers.
Following the approach of the section \ref{4.3}, we introduce the ensemble of
non-interacting particles, $\zeta_{ij}=0,1,2,...,$ indexed by $i,j\ge1.$
Again, the particles $\zeta_{ij}$ are not identical; the statistical weight
for the particle at the location $\left(  i,j\right)  $ to be in the state $k$
is equal to $\left(  x^{i+j-1}\right)  ^{k}.$ The volume $v\left(  \zeta
_{ij}\right)  $ associated to the particle $\zeta_{ij}$ when it is in the
state $k,$ is $\left(  i+j-1\right)  k.$ The weight of the configuration
$\mathbf{\zeta}$ of the particles $\zeta_{ij}$ is given by the same formula
(\ref{u30}).

Pushing the analogy with the section \ref{4.3} further, one should associate
with every configuration $\mathbf{\zeta}$ the collection of solid hooks: if
the particle $\zeta_{ij}$ is in the state $k,$ we have to take $k$ hooks of
the volume $\left(  i+j-1\right)  .$ One such hook is made from $\left(
i+j-1\right)  $ unit cubes; one can visualize it as having its vertex cube at
the location $\left(  i,j\right)  ,$ and going down and to the left till it
hits the coordinate axes. The temptation is now to build from every such
collection of hooks a skyscraper, so that different hook collections will be
transformed into different skyscrapers. This, however, seems to be impossible.
What is possible is somewhat more complicated construction. Namely, before
building a skyscraper from a given collection of hooks, one has to saw each
hook in two pieces along its bisecting plane. In this way we obtain twice as
many blocks as we had hooks, of half-integer volumes. It turns out that one
can then rearrange those blocks spatially, using only translations, in such a
way that the resulting shape will be a skyscraper, and the correspondence thus
obtained is one to one. This rearrangement is a special case of
Robinson-Schensted-Knuth correspondence, see \cite{Kn} or \cite{BKn}. It is
not at all straightforward, so we will not reproduce it here.

\section{Conclusion}

In this paper we have described the explicit geometric construction, which
predicts the asymptotic shape of some stat. mechanical and combinatorial
objects. It is worth mentioning that the method presented should work in
combinatorics whenever the underlying probability measure has certain locality
property, namely that the distant portions of the combinatorial object under
consideration are weakly dependent. This locality property is in fact the key
feature behind the results obtained in all the papers cited. It also holds for
the corresponding problems of statistical mechanics, like the validity of the
Wulff construction, and is crucial there as well.\medskip

\textbf{Acknowledgment.\ }I want to thank the colleagues with whom I was
discussing various problems which are discussed in this review, and in
particular A. Dembo, R. Kenyon, R. Kotecky, G. Olshanksi, A. Vershik and O. Zeituni.

\end{document}